\documentclass[prd,twocolumn,superscriptaddress,floatfix,amsmath,amssymb,amsfonts,nofootinbib]{revtex4-1}

\usepackage{float} 

\usepackage[normalem]{ulem}
\usepackage[english]{babel}
\usepackage{graphicx}
\usepackage{dcolumn}
\usepackage{bm}
\usepackage{blindtext}
\usepackage{verbatim}
\usepackage{relsize}
\usepackage{mathrsfs}
\usepackage{musicography}
\usepackage{amsmath}
\usepackage{blindtext}
\usepackage{cancel}
\usepackage{physics}
\usepackage{upgreek}
\usepackage{epstopdf}
\usepackage{mathtools}
\usepackage{tensor}
\usepackage{color}
\usepackage[usenames,dvipsnames]{pstricks}
\usepackage{epsfig}
\usepackage{pst-grad} 
\usepackage{pst-plot} 
\usepackage{hyperref}
\usepackage{verbatim}
\usepackage{slashed}

\renewcommand{\l}{\ell}
\newcommand{\m}{\mu}
\renewcommand{\d}{\sigma}
\newcommand{\w}{\upomega}

\newcommand{\mf}{\mathsf}

\newcommand{\ii}{\mathrm{i}}

\newcommand{\M}{\mathcal{M}}
\renewcommand{\L}{\mathcal{L}}
\renewcommand{\r}{\hat{\rho}}

\newcommand{\tc}[1]{\textsc{#1}}
\allowdisplaybreaks[1] 

\begin{document}

\title{Entanglement harvesting: detector gap and field mass optimization}

\author{H\'ector Maeso-Garc\'ia}
\email{hmaesoga@uwaterloo.ca}

\affiliation{Centre de Formació Interdisciplinària Superior (CFIS) -
Universitat Politècnica de Catalunya (UPC)}
\affiliation{Department of Applied Mathematics, University of Waterloo, Waterloo, Ontario, N2L 3G1, Canada}

\author{T. Rick Perche}
\email{trickperche@perimeterinstitute.ca}

\affiliation{Department of Applied Mathematics, University of Waterloo, Waterloo, Ontario, N2L 3G1, Canada}
\affiliation{Perimeter Institute for Theoretical Physics, Waterloo, Ontario, N2L 2Y5, Canada}
\affiliation{Institute for Quantum Computing, University of Waterloo, Waterloo, Ontario, N2L 3G1, Canada}

\author{Eduardo Mart\'{i}n-Mart\'{i}nez}
\email{emartinmartinez@uwaterloo.ca}

\affiliation{Department of Applied Mathematics, University of Waterloo, Waterloo, Ontario, N2L 3G1, Canada}
\affiliation{Perimeter Institute for Theoretical Physics, Waterloo, Ontario, N2L 2Y5, Canada}
\affiliation{Institute for Quantum Computing, University of Waterloo, Waterloo, Ontario, N2L 3G1, Canada}

\begin{abstract}
    We investigate the protocol of entanglement harvesting, where two spacelike separated particle detectors extract quantum correlations from a quantum field. Specifically, we analyze the role of the mass of the field and the energy gap of the detectors in the protocol. Perhaps surprisingly, we find that there are regimes in which the entanglement harvested can increase with the mass of the field by decreasing the noise experienced by the detectors. Finally, we study the optimal relationship between the gap of the detectors and the other parameters of the setting that maximizes the entanglement harvested, showing that a small mass can improve the protocol even in this case. 
\end{abstract}

\maketitle

\section{Introduction}

It is well known that Hadamard states of quantum fields can contain entanglement even between spacelike separated regions~\cite{vacuumBell,vacuumEntanglement}. The presence of entanglement in the field is associated with many fundamental phenomena from holography to Hawking radiation and the black hole information loss problem~\cite{Preskill1992,HawkingLoss, BlackHoles1, BlackHoles2, BlackHoles3}. However quantifying the entanglement present in quantum fields remains an elusive task. In fact, we only have a handful of techniques that can only be used in specific scenarios~\cite{witten}.

Nevertheless, we have tools to compute the amount of entanglement displayed by the field between any pair of regions of spacetime that can be accessed by physically measurable probes. Namely, the entanglement displayed by the field between these regions can be extracted by particle detectors in a setup that has been commonly known as \emph{entanglement harvesting}~\cite{Valentini1991,reznik1,Pozas-Kerstjens:2015}. The entanglement harvesting protocol considers initially uncorrelated probes that couple locally to the quantum field in order to extract entanglement from it.

In this manuscript we study in detail the effect that different parameters have in the entanglement harvesting protocol, with special focus on the field's mass and the detectors' energy gap. It is a known fact that the correlations of a quantum field decay exponentially with the mass of the field. For this reason, one might have expected that the entanglement harvested from particle detectors would also also decay with mass. However, the entanglement harvested by local probes is a competition between non-local correlations of the field and local noise terms associated with the detectors' excitation probability. Despite the fact that the field correlations decay exponentially with its mass, we find that there are regimes where the local noise terms decay even faster. For this reason, while particle detectors capture the exponential decay of correlations for large enough mass, there are situations in which having a small field mass allows us to harvest more entanglement than we would in identical setups for a massless field. What is more, there are scenarios where a pair of detectors cannot harvest entanglement from massless fields, but by adding a small field mass the very same detectors are able to harvest entanglement.

We also study the optimal regimes where the parameters of smoothly localized  detectors are tuned to harvest the maximum possible amount of entanglement. We find simple approximate relations between the parameters of the setup that optimize the protocol. We also find that, even in this optimal case, a small finite mass of the quantum field enhances the entanglement harvesting protocol. This result implies that although the entanglement present in a quantum field decreases with its mass, the entanglement that can be accessed by physical probes does not share this monotonic behaviour in all regimes.

This manuscript is organized as follows. In Section \ref{sec:KG} we review the basic properties of a massive real scalar quantum field, paying special attention to the effect that the mass has on the field correlations. In Section \ref{UDW} we review the Unruh-DeWitt (UDW) particle detector model and the setup of entanglement harvesting. In Section \ref{protocol}, we describe an explicit protocol with specific detector shapes, an discuss its main features. In Section \ref{mass} we study the behaviour of the accessible entanglement in the quantum field as a function of the field's mass. In Section \ref{sec:optimizing} we find the parameters which optimize the protocol of entanglement harvesting. In Section \ref{conclusions} we summarize the conclusions of our work.

\section{A Massive Klein-Gordon field}\label{sec:KG}

Consider a $D = n+1$ dimensional spacetime $M$ and a real scalar field $\phi:M\longrightarrow \mathbb{R}$ whose dynamics are determined by the minimally coupled Klein-Gordon equation
\begin{equation}\label{KG}
    (\nabla_\mu \nabla^\mu -m^2)\phi = 0,
\end{equation}
where $\nabla$ denotes the metric compatible torsion free connection and $m$ is a constant with units of energy, which is commonly referred to as the field's mass for reasons that will be discussed below. If the spacetime $M$ is globally hyperbolic, Eq. \eqref{KG} admits a unique solution for initial conditions given in any Cauchy surface. It is then possible to find an orthonormal\footnote{Orthonormal here refers to the Klein-Gordon inner product:
\begin{equation}
    (\phi_1,\phi_2) = \ii \int_{\Sigma} \dd \Sigma^\mu \left(\phi_1^* \nabla_\mu \phi_2 - \nabla_\mu \phi_1^* \,\phi_2\right),
\end{equation}
where $\Sigma$ is a Cauchy surface and $\dd \Sigma^\mu$ denotes its volume element with the unit normal.
} basis of solutions $\{u_{\bm k}(\mf x),u^*_{\bm k}(\mf x)\}_{\bm k}$. Let us assume that the labels $\bm k$ are a continuous set $\bm k\in \mathbb{R}^{n}$, as this is the case for many different spacetime backgrounds, such as Minkowski spacetime. In terms of this basis, any classical solution to the Klein-Gordon equation can be written as
\begin{equation}
    \phi(\mf x) = \int \dd^n \bm k \left(u_{\bm k} ( \mf x) {a}_{\bm k} + u_{\bm k}^* ( \mf x) {a}^*_{\bm k}\right),
\end{equation}
where the coefficients $a_{\bm k}$ are determined by the initial conditions.

In order to canonically quantize the Klein-Gordon field, we promote the coefficients $a^*_{\bm k}$ and $a_{\bm k}$ to the creation and annihilation operators, $\hat{a}_{\bm k}^\dagger$ and $\hat{a}_{\bm k}$. By imposing the commutation relations
\begin{equation}
    \comm{\hat{a}^{\vphantom{\dagger}}_{\bm k}}{\hat{a}^\dagger_{\bm k'}} = \delta^{(3)}(\bm k - \bm k')
\end{equation}
we ensure that the so-defined field operator $\hat{\phi}(\mf x)$ and its conjugate momentum satisfy canonical commutation relations. The creation and annihilation operators then define the vacuum state $\ket{0}$ via $\hat{a}_{\bm k} \ket{0} = 0\:\:\forall \bm k$, and the Fock space is constructed by repeated applications of the creation operators $\hat{a}^\dagger_{\bm k}$ on $\ket{0}$.

\subsection{A Klein-Gordon field in Minkowski spacetime}\label{sub:KGmink}

In this paper, we will focus on a free Klein-Gordon field in Minkowski spacetime. In this case, there are simple interpretations for the mass of the field, and it is possible to obtain closed form expressions for all the correlation functions of the field in the vacuum state. In Minkowski spacetime, a natural choice for orthonormal basis of solutions to the Klein-Gordon equation is the plane-wave basis
\begin{equation}\label{modes}
    u_{\bm k}(\mf x) = \frac{1}{(2\pi)^{\frac{n}{2}}} \frac{e^{\ii \mf k \cdot \mf x}}{\sqrt{2 \omega_{\bm k}}},
\end{equation}
where $\omega_{\bm k} = \sqrt{m^2 + \bm k^2}$ and $\mf k \cdot \mf x = - \omega_{\bm k}\, t + \bm k \cdot \bm x$ with $\mf x = (t,\bm x)$ in inertial coordinates. The explicit dependence $\omega_{\bm k}(\bm{k})$ then defines a dispersion relation, so that the group velocity of the mode with momentum $\bm k$ is given by $\dv{\omega_{\bm k}}{\bm k} = \bm k/\omega_{\bm k} < 1$. In this sense, the mass acts as an ``inertia'' for the field, reducing the speed of propagation of information in spacetime. 

The parameter $m$ can also be associated with the mass of the quantum field in the following sense: from the modes of Eq. \eqref{modes}, it is possible to show that the normal ordered Hamiltonian associated to the space slices \mbox{$t=\text{const.}$} reads
\begin{equation}
    :\!\!\:\!\hat{H}\!\!:\, = \int \dd^n \bm k \,\omega_{\bm k} \hat{a}_{\bm k}^\dagger \hat{a}_{\bm k}.
\end{equation}
In particular, the smallest value of energy that can be acquired by a field excitation happens with the Fock state $\hat{a}_{\bm 0}^\dagger \ket{0}$, with energy $m$. This corresponds to a Fock excitation with zero momentum. The fact that this is the smallest energy excitation admissible by this quantum field theory allows one to interpret $m$ as the rest mass of a `particle' excitation, and thus, with the mass of the field.

\begin{figure}[b!]
    \includegraphics[scale=0.73]{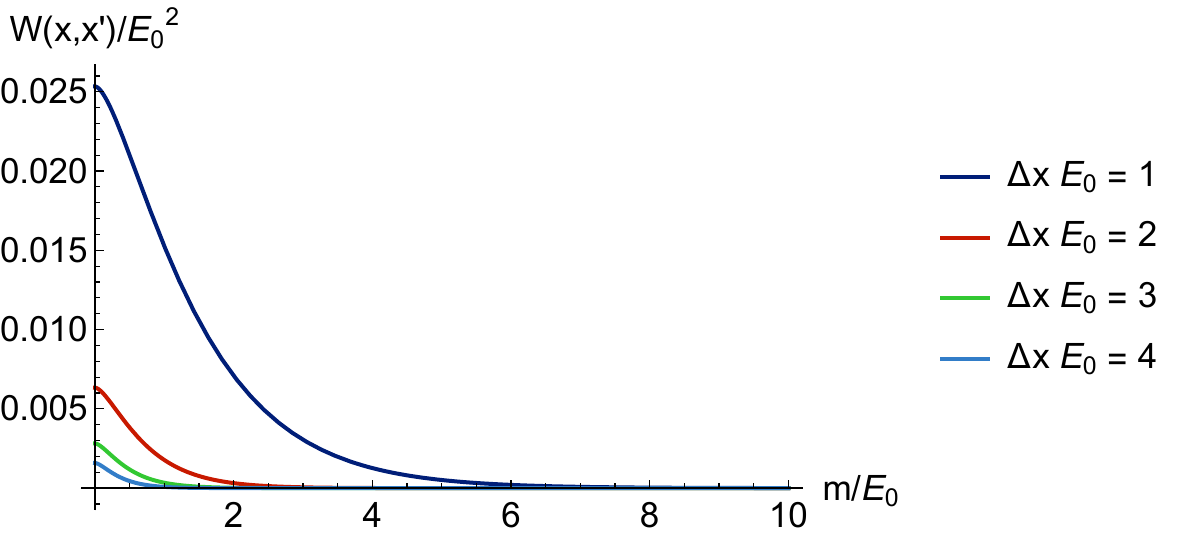}
    \caption{Correlation function of a Klein-Gordon field in $3+1$ dimensional Minkowski spacetime as a function of its mass for spacelike separated events (separated by a proper distance $\Delta\mf{x}$) in terms of an arbitrary energy scale $E_0$. }\label{fig:scalar}
\end{figure}

The mass of the quantum field can also be interpreted as the parameter that controls the decay of correlations within the quantum field. In fact, the only dimensionless parameter that can be built from the spacetime separation between events $\mf x$ and $\mf x'$  (for a massive field in flat spacetimes) is $m \Delta \mf x$, where \mbox{$\Delta\mf x^2 = \eta_{\mu\nu} (x-x')^\mu (x-x')^\nu$}. In particular, it is possible to show that the Wightman two-point function of the field of the vacuum, \mbox{$W(\mf x,\mf x') = \bra{0}\!\hat{\phi}(\mf x) \hat{\phi}(\mf x')\!\ket{0}$}, can be written as $f(m \Delta \mf x)/(\Delta \mf x)^{D-2}$, where $f$ is a dimensionless function which is regular in the limit $\Delta \mf x \rightarrow 0$. In fact, we can write the (regularized) two-point function explicitly as
\begin{equation} \label{eq:wightman}
  W(\mf x,\mf x')=\frac{2}{(4\pi)^\frac{n}{2}} \left(\frac{4m^2}{\Delta \mf x_\epsilon }\right)^{\frac{n-2}{4}}
   K_{\frac{n}{2}-1}\left( m \Delta \mf x_\epsilon \right),
\end{equation}
where $K_n$ denotes the modified Bessel functions of the second kind and $\Delta \mf x_\epsilon$ denotes the regularized spacetime separation,
\begin{equation}
    \Delta \mf x_\epsilon^2 = -(t-t'-\ii \epsilon)^2 + (\bm x - \bm x')^2.
\end{equation}
The Wightman function is a distribution that can be thought of as the limit of $\epsilon\to 0$ of the regularized expression above. To work with the Wightman function it is convenient to use the regularized version to deal with the coincidence limit $\Delta\mf x \to 0$. In Fig \ref{fig:scalar}, we plot the behaviour of the Wightman function with the mass of the field for fixed positive values of the invariant spacelike interval in terms of an arbitrary fiducial energy scale $E_0$. In the plot, we see the exponential decay of the correlations with the mass of the field. 

\section{Entanglement harvesting and the UDW Model}\label{UDW}

This section has the purpose of reviewing the entanglement harvesting protocol. We begin by introducing the well-known Unruh-DeWitt particle detector model and then review its application to entanglement harvesting.





\subsection{The UDW model}\label{sub:UDW}

There are different ways to access the information encoded in a quantum field. A common approach consists of using particle detector models to locally probe the field. In this context, a particle detector is a localized non-relativistic quantum system with an internal degree of freedom that can couple locally to a quantum field while preserving the causality and covariance of the theory\footnote{For pointlike detectors the preservation is exact~\cite{martin-martinez2015,us2}. For spatially smeared detectors the preservation is approximate within the limits of applicability of the model~\cite{martin-martinez2015,us2,pipoFTL}}. 

Among the simplest and most successful models for particle detectors is the Unruh-DeWitt (UDW) model \cite{Unruh1976,DeWitt}. This model has been extensively used to study a wide variety of phenomena such as the Unruh~\cite{Unruh1976,Unruh-Wald,takagi,matsasUnruh,mine} and Hawking effects~\cite{Unruh1976,HawkingRadiation,bhDetectors}, quantum energy teleportation~\cite{teleportation, teleportation2014}, modelling quantum and classical communication in relativistic setups~\cite{Jonsson2,martin-martinez2015,Jonsson3,Jonsson4,Katja,Landulfo,Casals,KojiCapacity,EricksonCom}, as well as to approach more fundamental aspects of QFT such as defining a measurement theory for quantum fields~\cite{chicken}. Albeit simple, this interaction captures the fundamental features of common experimental setups in quantum optics~\cite{richard} and high-energy physics~\cite{neutrinos,antiparticles,carol}.




We now review the simplest version of the UDW model. The detector is modelled by a two-level quantum system to which we associate a timelike trajectory $\mf z(\tau)$ and that couples locally to a real scalar field $\hat{\phi}(\mf{x})$. Here $\tau$ denotes the proper time of the trajectory $\mf z(\tau)$. We denote the proper energy gap of the detector by $\Omega$ and its ground and excited states by $\ket{g}$ and $\ket{e}$. The detector's free Hamiltonian that generates time evolution with respect to $\tau$ is prescribed as 
\begin{equation}
    \hat{H}_{\text{d}} = \Omega \hat{\sigma}^+ \hat{\sigma}^-,
\end{equation}
where $\hat{\sigma}^+ = \ket{e} \!\! \bra{g}$ and $\hat{\sigma}^- = \ket{g} \!\! \bra{e}$ are the ladder operators of the two-level system. The coupling between the detector and the background field is modelled in the interaction picture using the Hamiltonian weight \cite{us}
\begin{equation}
    \hat{h}_I(\mf x) = \lambda \Lambda(\mf x) \hat{\mu} (\tau) \hat{\phi}(\mf x),
\end{equation}
where $\lambda$ is the coupling strength, $\Lambda(\mf x)$ is the spacetime smearing function, which defines the region of spacetime where the interaction takes place, and
\begin{equation}
    \hat{\mu} (\tau) = e^{\ii \Omega \tau}\hat{\sigma}^+  + e^{-\ii \Omega \tau}\hat{\sigma}^- 
\end{equation}
is the monopole moment of the detector.

The joint state of the detector-field system evolves according to the time evolution operator
\begin{equation} \label{eq:TEO}
    \hat{U} = \mathcal{T} \text{exp}\left(-\ii \int \dd V \hat{h}_I(\mf x) \right).
\end{equation}
Here, $\dd V$ is the invariant volume element of spacetime and $\mathcal{T}\text{exp}$ denotes the time ordered exponential. We remark that the expression above is in principle dependent on the choice of the time parameter used to define the time ordering. However, in~\cite{us2} it is shown that this dependence on the time parameter choice only affects the final state of the detectors system in specific cases. In particular, for the applications present in this manuscript, it was shown that to leading order in $\lambda$, the detectors state is independent of the time parameter chosen to prescribe $\hat{U}$, therefore ensuring the covariance of the model. We will assume that, before the interaction, the joint detector-field density operator $\hat{\rho}^{(0)}$ is in an uncorrelated state
\begin{equation}
    \hat{\rho}^{(0)} = \hat{\rho}_{\phi}^{(0)} \otimes \hat{\rho}_{\textsc{d}}^{(0)},
\end{equation}
 where $\hat{\rho}_{\phi}^{(0)}$ is the initial state of the field and $\hat{\rho}_{\textsc{d}}^{(0)}$ the initial state of the detector. The time-evolved state due to the interaction between the detector and the field is given by 
\begin{equation}\label{eq:EQUATIONALWAYS}
    \hat{\rho} = \hat{U} \hat{\rho}^{(0)} \hat{U}^{\dagger}.
\end{equation}
Recall that the detector is used as a probe to extract information from the field. Thus, once the interaction is switched off, the degrees of freedom corresponding to the field state no longer affect the detector. The detector state is obtained by tracing out the field's degrees of freedom, so that the final state of the detector is given by
\begin{equation}
    \hat{\rho}_{\textsc{d}} = \text{Tr}_{\phi}(\hat{U} \hat{\rho}^{(0)} \hat{U}^{\dagger}).
\end{equation}
$\hat{\rho}_{\textsc{d}}$ can be completely determined by $\hat{\rho}_\textsc{d}^{(0)}$ and the field's $n$-point functions. In particular, if the field's state $\hat{\rho}_\phi^{(0)}$ is a zero-mean Gaussian state (such as the vacuum state), then $\hat{\rho}_{\textsc{d}}$ is entirely determined by the field's two-point function $W_{\hat{\rho}_\phi}(\mf x,\mf x') = \tr\left(\hat{\rho}_\phi \hat{\phi}(\mf x) \hat{\phi}(\mf x')\right)$.

\subsection{The Entanglement Harvesting Protocol}\label{sub:harvest}

Here we review a simple protocol that allows two spacelike separated particle detectors to extract entanglement from a quantum field: the \emph{entanglement harvesting} protocol. This setup has exhaustively
been explored in the literature, with most studies mainly focusing on the entanglement properties of massless fields. In this manuscript we will study in detail the behaviour of entanglement with the mass of the field. Consider a pair of particle detectors $\textsc{A}$ and $\textsc{B}$ initially in their ground states $\ket{g_{\textsc{a}}} \!\! \bra{g_{\textsc{a}}}$,$  \ket{g_{\textsc{b}}}\!\!\bra{g_{\textsc{b}}}$. We couple these particle detectors to a real scalar field that is in its vacuum state $\ket{0}$ prior to the interaction. The initial state of the joint detectors-field system is 
\begin{equation}
    \r^{(0)} = \ket{g_{\textsc{a}}} \!\! \bra{g_{\textsc{a}}} \otimes  \ket{g_{\textsc{b}}}\!\!\bra{g_{\textsc{b}}} \otimes \ket{0}\!\!\bra{0}.
\end{equation}
The Hamiltonian weight for the interaction of the detectors with the field is
\begin{equation}
     \hat{h}_I(\mf x) = \sum_{i \in \{\textsc{a}, \textsc{b} \}} \lambda_{i} \Lambda_{i}(\mf x) \hat{\mu}_{i} (\tau_{i}) \hat{\phi}(\mf x).
\end{equation}
The final state for the joint detectors-field system is obtained via the time-evolution operator using Eq. \eqref{eq:EQUATIONALWAYS} with $\hat{U}$ given by Eq. \eqref{eq:TEO}. 

We will assume that the detectors are weakly coupled to the field and proceed perturbatively on their coupling strengths, assuming that both $\lambda_{\textsc{a}}$ and $\lambda_{\textsc{b}}$ are of the same order of magnitude. The Dyson expansion for the time evolution operator reads 
\begin{equation}
    \hat{U} = \openone + \hat{U}^{(1)} + \hat{U}^{(2)} + \mathcal{O}(\lambda^3),
\end{equation}
where
\begin{align}
    U^{(1)} &= -\ii \int \dd V  \hat{h}_I(\mf x), \\
    U^{(2)} &= - \int \dd V  \int \dd V' \hat{h}_I(\mf x)  \hat{h}_I(\mf x') \theta{(t-t')},
\end{align}
where we have chosen an arbitrary time coordinate $t$ for the time ordering. The notation $\mathcal{O}(\lambda^k)$ refers to products of $\lambda_{\textsc{a}}$ and $\lambda_{\textsc{b}}$ of order $k$.
Notice that the term $\hat{U}^{(k)}$ is of order $\mathcal{O}(\lambda^k)$. This expansion allows us to express the time-evolved final state of the detectors-field system after the interaction as
\begin{equation}
    \hat{\rho} =  \hat{\rho}^{(0)} +  \hat{\rho}^{(1)} +  \hat{\rho}^{(2)} + \mathcal{O}(\lambda^3), 
\end{equation}
where
\begin{align}
    \r^{(1)} &= \hat{U}^{(1)} \r^{(0)}  +  \r^{(0)} \hat{U}^{(1) \, \dagger},  \\
    \r^{(2)} &= \hat{U}^{(2)} \r^{(0)}  + \hat{U}^{(1)} \r^{(0)} \hat{U}^{(1) \, \dagger}  +  \r^{(0)} \hat{U}^{(2) \, \dagger}.
\end{align}

We are interested in studying under which conditions this interaction allows the detectors to extract entanglement from the field. We trace out the field to obtain the final state of both detectors, $\hat{\rho}_{\textsc{ab}} = \text{Tr}_{\phi}(\r)$. In the basis $\{\ket{g_\textsc{a} g_{\textsc{b}}}, \ket{g_\textsc{a} e_{\textsc{b}}}, \ket{e_\textsc{a} g_{\textsc{b}}}, \ket{e_\textsc{a} e_{\textsc{b}}}\}$, the detectors final density operator is represented by the matrix
\begin{equation}\label{eq:hectorLearnsToCiteEq}
    \hat{\rho}_{\textsc{ab}} = \left(
\begin{array}{cccc}
1-\mathcal{L}_\textsc{aa}-\mathcal{L}_\textsc{bb} & 0 & 0 & \mathcal{M}^\ast \\
0 & \mathcal{L}_\textsc{bb} & \mathcal{L}_\textsc{ab}^\ast & 0 \\
0 & \mathcal{L}_\textsc{ab} & \mathcal{L}_\textsc{aa} & 0 \\
\mathcal{M} & 0 & 0 & 0 \\
\end{array}\right)+O(\lambda^3) \;,
\end{equation}
 with
\begin{align} \label{eq:L,M}
    \L_{ij} &=\lambda_i \lambda_j \int \dd V \dd V' \Lambda_{i}(\mathsf{x}) \Lambda_{j}(\mathsf{x}') e^{-\ii(\Omega_{i} \tau_i - \Omega_{j} \tau_j^{'})}  W(\mf x,\mf x'), \\
    \M &= -\lambda_{\textsc{a}}\lambda_{\textsc{b}}\int \dd V \dd V' \Lambda_{\textsc{a}}(\mathsf{x}) \Lambda_{\textsc{b}}(\mathsf{x}') e^{\ii(\Omega_{\textsc{a}} \tau_\tc{a} + \Omega_{\textsc{b}} \tau_\textsc{b}^{'})}\nonumber\\& \:\:\:\:\:\:\:\:\:\:\:\:\times  \big(W(\mf x,\mf x') \theta(t-t') + W(\mf x',\mf x) \theta(t'-t) \big).
\end{align}
for $i,j$ in $\{\textsc{A}, \textsc{B} \}$. 

In order to quantify the entanglement acquired by the detectors we use the negativity, which is a trustworthy entanglement monotone for two two-dimensional quantum systems~\cite{VidalNegativity}. Although both negativity and concurrence are popular for harvesting with two-level UDW detectors, the negativity has the advantage that it is well defined and easy to compute also for higher dimensional quantum systems and helps comparing the results to more general scenarios. The negativity of a bipartite state $\r_{\textsc{ab}}$  is defined as the absolute sum of the negative eigenvalues of the partial transpose of $\r_{\textsc{ab}}$, with respect to either $\textsc{A}$ ($\r_{\textsc{ab}}^{t_\textsc{a}}$) or $\textsc{B}$ ($\r_{\textsc{ab}}^{t_\textsc{b}}$). Namely, the negativity is given by
\begin{align}
    \text{max}\left(\sum_{\nu_i < 0} \left|\nu_i\right|\:,\:\:0\:\:\right),
\end{align}
where $\nu_i$ are the eigenvalues of $\r_{\textsc{ab}}^{t_{\textsc{a}}}$ (which coincide with the eigenvalues of $\r_{\textsc{ab}}^{t_{\textsc{b}}}$). At order $\mathcal{O}(\lambda^2)$ the negativity takes the form
\begin{equation} \label{eq:negativeEigenvalue}
    \mathcal{N} = \text{max}\left(0,\sqrt{|\mathcal{M}|^2+\frac{(\mathcal{L}_{\textsc{aa}} - \mathcal{L}_{\textsc{bb}})^2}{4}}-\frac{\mathcal{L}_{\textsc{aa}} + \mathcal{L}_{\textsc{bb}}}{2} \right).
\end{equation}

 Observe that at leading order the negativity is determined by $\mathcal{M}$, $\mathcal{L}_{\textsc{aa}}$ and $\mathcal{L}_{\textsc{bb}}$. Out of these terms, $\mathcal{M}$ is the only one which contains non-local information involving both detectors. The terms  $\mathcal{L}_{\textsc{aa}}$ and $\mathcal{L}_{\textsc{bb}}$ are the local terms, namely the excitation probabilities for each detector.
 One can easily see that if the detectors are identical, $\mathcal{L}_{\textsc{aa}} = \mathcal{L}_{\textsc{bb}} = \mathcal{L}$, the negativity simplifies to
 \begin{equation}\label{eq:simpleNeg}
    \mathcal{N}(\hat{\rho})  = \max(0, |\mathcal{M}| - \mathcal{L}).
\end{equation}
and entanglement appears when the correlation term $\mathcal{M}$ `wins' over the local noise terms $\mathcal{L}$\footnote{{In fact the argument that entanglement at leading order is always a competition between local noise and the correlation term $\mathcal{M}$  can be made even if the detectors are not identical since we can bound $\mathcal{N}$ using
 \begin{align*}
    &\sqrt{|\mathcal{M}|^2+\frac{(\mathcal{L}_{\textsc{aa}} - \mathcal{L}_{\textsc{bb}})^2}{4}} - \frac{\mathcal{L}_{\textsc{aa}} + \mathcal{L}_{\textsc{bb}}}{2} \nonumber \\* 
     &\leq |\mathcal{M}|+\frac{|\mathcal{L}_{\textsc{aa}} - \mathcal{L}_{\textsc{bb}}|}{{2}} - \frac{\mathcal{L}_{\textsc{aa}} + \mathcal{L}_{\textsc{bb}}}{2} \leq |\mathcal{M}|,
\end{align*}
so that we obtain $\mathcal{N} \leq |\mathcal{M}|$}.
}



There are two ways in which the detectors can get entangled. On the one hand two detectors whose interaction regions are causally connected can exchange information through the quantum field, and through that communication they can get correlated. On the other hand, even two detectors that remain spacelike separated can get entangled through their interactions with the field. As mentioned in the introduction, this is possible because the vacuum of the quantum field contains entanglement between spacelike separated regions\cite{vacuumEntanglement,vacuumBell}. Only the genuine extraction of entanglement from the field that is not mediated by communication should be considered as entanglement harvesting~\cite{ericksonNew}. There are scenarios where one can possibly have both mechanisms at work: one could be harvesting timelike or lightlike correlations from the vacuum and yet acquiring some extra entanglement due to communication. In those cases it is important to distinguish the two different mechanisms and their respective contributions to harvesting. 
The contribution of each of these two mechanisms was analyzed in detail in~\cite{ericksonNew}. Namely, the expression for $\mathcal{M}$ involves the two-point correlator $W(\mf x,\mf x')$ sampled at the two different regions of interaction, given by the supports of $\Lambda_{\textsc{a}}(\mathsf{x})$ and $\Lambda_{\textsc{b}}(\mathsf{x})$. The Wightman function can be then separated into its real and imaginary parts. The imaginary part only depends on the field commutator $[\hat{\phi}(\mf x), \hat{\phi}(\mf x')]$ which is independent of the field's state and mediates communication~\cite{Jonsson2,martin-martinez2015,Jonsson3}. In flat spacetime, and at leading order, the contribution of $\Im \mathcal{M}$ can be interpreted as communication between the detectors. The real part of $W(\mf x,\mf x')$ only depends on the (state dependent) field anti-commutator $\{\hat{\phi}(\mf x), \hat{\phi}(\mf x')\}$ and it can be interpreted as the contribution of the field state to the field's correlator.
 
 Because of this, in~\cite{ericksonNew} it is suggested that anytime that there is any causal contact between detectors, one can estimate if the acquired entanglement is harvested (as opposed to generated by communication). This can be done by quantifying the contributions to negativity coming from the real part of the Wightman function and the imaginary one. If the imaginary part dominates, the entanglement generated between the detectors is not harvested. We will use this to see when we have entanglement harvesting in the different regimes we analyze.

As it is well-known, the two-point correlator decays with the spacetime separation. This implies that the larger the distance between the interaction regions, the more suppressed the term $\mathcal{M}$ will be. Since the noise $\mathcal{L}$ experienced by the detectors is local, it is clear that entanglement harvesting decreases with distance. This trend is well-known to hold both in flat and curved spacetimes.

Entanglement harvesting also depends on the internal structure of the detectors, which in this case is defined by the parameters $\Omega_\textsc{a}$ and $\Omega_{\textsc{b}}$. For inertial co-moving detectors, it is commonly believed that using detectors with different energy gaps decreases the negativity. In Appendix~\ref{app:Omegas} we show that this is the case. Therefore it is very common to assume that the energy gap of both detectors is the same, namely $\Omega_{\textsc{a}} = \Omega_{\textsc{b}} = \Omega$, and we will work under this assumption in this manuscript. 

Finally, the behaviour of entanglement harvesting with the field's mass has not been studied in detail in previous literature. This is mostly due to the fact that until recently~\cite{neutrinos}, the main physical process that was modelled with particle detectors was the light-matter interaction (which involves a massless field). Another reason why  massive fields have been given less attention in the past is the fact that the field correlations decay with its mass (see Fig. \ref{fig:scalar}). This could naturally lead one to the intuition that harvesting from increasingly massive fields results in less entangled detectors. One of the goals of this manuscript is to study the effects that the field's mass has in the entanglement harvesting protocol and to reveal some unexpected subtleties in its behaviour.



\section{Harvesting spacelike entanglement with Gaussian smearings}\label{protocol}  

For concreteness in our study, in this section we analyze entanglement harvesting using Gaussian smearing functions. In order to obtain explicit results, we focus on the concrete example of two inertial comoving identical UDW detectors in (3+1) dimensional Minkowski spacetime. The detectors couple to the vacuum of a massive scalar quantum field and have Gaussian smearing and switching functions prescribed in their comoving frame. 

Let us choose the quantization frame $(t,\bm x)$ to be comoving with the two detectors' centre of mass trajectories. As described in Section~\ref{sub:UDW} the detectors move along trajectories $\mf{z}_{\textsc{a}}(t) = (t,\bm 0)$ and $\mf{z}_\textsc{b}(t) = (t,\bm L)$. Under the usual Fermi-Walker rigidity condition~(see, e.g.~\cite{us,mine}), we prescribe their spacetime smearing functions as
\begin{align}
    \Lambda_{\textsc{a}}(\mf x) &= \chi(t)F(\bm x),\label{lambdaA}\\
    \Lambda_{\textsc{b}}(\mf x) &= \chi(t)F(\bm x-\bm L),\label{lambdaB}
\end{align}
and assume their energy gaps and coupling constants to be identical, $\Omega_\textsc{a} = \Omega_\textsc{b} = \Omega$. The functions $F(\bm x)$ and $\chi(t)$ are chosen as the following Gaussians
\begin{align}
    \chi(t) &= \frac{1}{\sqrt{2\pi}}e^{-\frac{t^2}{2T^2}},\label{chi}\\
    F(\bm x) &= \frac{1}{(2\pi R^2)^{3/2}}e^{-\frac{\bm x^2}{2R^2}}.\label{F}
\end{align}
$T$ controls the time duration of the interaction, 
and $F(\bm x)$ defines a probability distribution with standard deviation $R$ which controls the size of the detector. 

With the choices of Eqs. \eqref{chi} and \eqref{F}, we obtain $\mathcal{L}_{\textsc{aa}} = \mathcal{L}_{\textsc{bb}}\equiv\mathcal{L}$, so that we are in the regime where the negativity is given by Eq. \eqref{eq:simpleNeg}. Moreover,  the terms $\mathcal{L}$ and $\mathcal{M}$ can be cast as a single momentum integral:
\begin{align} 
    \mathcal{L} &= \frac{\lambda^2T^2}{2 \pi^2}\int\frac{\dd |\bm k|}{2 \omega_{\bm k}} \, |\bm k|^2 \,e^{-|\bm k|^2 R^2}\, e^{-(\Omega+ \omega_{\bm k})^2 T^2} ,\\
    \mathcal{M} &= -\frac{\lambda^2T^2}{2 \pi^2}\int\frac{\dd |\bm k|}{2\omega_{\bm k}} \, |\bm k|^2 \,e^{-|\bm k|^2 R^2}\, e^{-(\Omega^2+ \omega_{\bm k}^2) T^2} \text{sinc}(|\bm k| |\bm  L|) \nonumber \\
    &\phantom{==============}\times\left(1- \text{erf}\left( \ii T \omega_{\bm k} \right) \right).
\end{align}


We are going to use the duration $T$ to set the rest of the scales of the setup. Correspondingly, we define dimensionless variables $\kappa = |\bm{k}|T$, $\w \coloneqq \Omega T$, $\mu \coloneqq m T$, $\sigma \coloneqq R / T$ and $\l \coloneqq |\bm{L}|/T$. Then, the interaction time $T$ sets a scale that can be used to compare all the magnitudes involved in the problem. In terms of the new dimensionless variables,
\begin{align} 
    \mathcal{L} &= \frac{\lambda^2}{2 \pi^2}\int\frac{\dd \kappa}{2 \omega_{ \kappa}} \, \kappa^2 \,e^{-\kappa^2 \d^2}\, e^{-(\w+ \omega_{\kappa})^2}\label{eq:Lk} ,\\
    \mathcal{M} &= -\frac{\lambda^2}{2 \pi^2}\int\frac{\dd \kappa}{2\omega_{\kappa}} \, \kappa^2 \,e^{-\kappa^2 \d^2}\, e^{-(\w^2+ \omega_{\kappa}^2)} \text{sinc}(\kappa \l) \label{eq:Mk} \\
    &\phantom{==============}\times\left(1- \text{erf}\left( \ii  \omega_{\kappa} \right) \right),\nonumber
\end{align}
where we have also adimensionalized  $\omega_{\kappa} := \omega_{\bm{k}}T $.

\begin{figure}[h]
    \includegraphics[width=8.5cm]{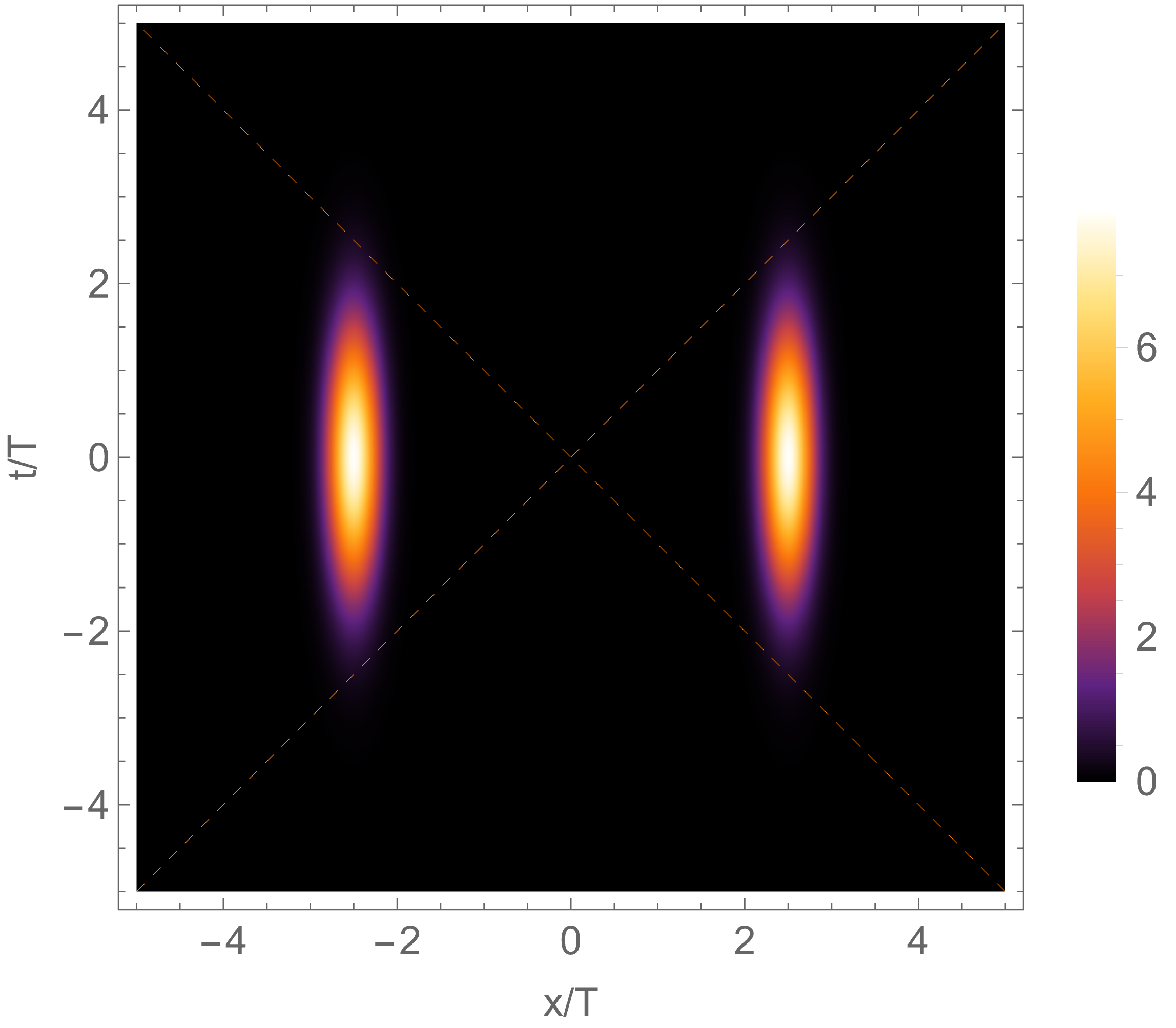}
    \caption{Schematic representation of smearing of two detectors separated by distance $\l = 5$ with smearing $\sigma = 0.2$. The dashed lines denote the $x-\ell/2 = \pm t$ lines.}\label{scheme}
\end{figure}

In order to claim that the entanglement acquired by the detectors is extracted from the quantum field, and not due to communication between the detectors, we consider regimes where the spacetime supports of the detectors, $\Lambda_\textsc{a}(\mf x)$ and $\Lambda_\textsc{b}(\mf x)$, are approximately spacelike separated. We note that due to the fact that both spacetime smearing functions are Gaussians, their tails always overlap. However, if $|\bm L|$ is large enough compared to $T$ and $R$ (i.e., $\l\gg 1$ and $\l\gg\sigma$), one would expect that to all intent and purposes the detectors are approximately spacelike separated (see, e.g.,\cite{martin-martinez2015}). In Fig. \ref{scheme} we show a spacetime diagram of the region of interaction of the two detectors, where we see the effective spacelike separation in practice. In our examples, we will keep $\l\geq 5$ and $\d\leq 0.4$, which are choices that ensure that the spacetime smearing functions are enough approximately spacelike separated so that the entanglement the detectors acquire is dominated by spacelike harvesting. This is a non-trivial requirement and, as we will see below, the field modes that the detector couples to can also determine whether their interaction can be considered to harvest spacelike entanglement or not. The modes the detectors are sensitive to  are not only determined by the detectors' size, but also by their energy gap and the field's mass. 

\subsubsection*{Spacelike separation and the detectors' gap}\label{spacelike}

Let us turn our attention to the conditions that ensure that the interactions of the detectors with the field prescribed above can be considered approximately spacelike separated for entanglement harvesting purposes. Fig. \ref{scheme} shows that for $\ell = 5$ the spacetime smearing functions can be considered to be approximately spacelike separated. In fact, for $\ell = 5$ and $\d = 0.2$ it can be shown that the integral of the product of the spacetime smearing functions over all spacetime is of the order\footnote{The most conservative estimate of the wellness of the approximate spacelike separation between the smearings is can be computed by propagating the full spacetime smearing function of one detector in the null direction that points towards the other until we get the maximum possible overlap, then compute the integral of their product. This gives a value still very small: $1.5 \times 10^{-4}$. } of $10^{-68}$.

However, the fact that the interaction regions are approximately spacelike separated might not be enough to ensure that the causal contact between the detectors (which are not compactly supported), is not responsible for most of the entanglement they acquire. In other words, we want to guarantee that the signalling between the Gaussian tails of the detectors' spacetime smearing (or even the small spatial overlap between their smearing functions) does not play any significant role in the entanglement they acquire.

For non-compact detectors that are approximately spacelike separated, the energy gap also plays a role on whether entanglement acquired between the detectors is genuinely harvested. Intuitively, due to resonance,  the detectors are most favoured to interact with field modes  whose frequency is of the order of the energy gap $\omega_{\bm k} \approx \Omega$ (i.e., $\omega_\kappa \approx \w$)\footnote{It is important to keep in mind that modes that are `most seen' by the detectors are not only regulated by the detector gap and resonance effects: the detectors' size and interaction time matter as well. The intuition is clearer for interactions that are `long enough' so that the Fourier transforms of the switching and smearing functions in the mode integrals \eqref{eq:Lk} and \eqref{eq:Mk} do not suppress the resonance effects at $\omega_\kappa \approx \w$.}, so that the wavelength of these field modes is given approximately by $1/\Omega$. That is, one would expect that if the detector gap is too small, then field modes with large wavelengths that are resonant with the detector gaps dominate any communication between the detectors. If the resonant field modes have large wavelengths as compared to the separation of the detectors, they may favour signalling effects even if the smearings are considered to be themselves approximately spacetime separated\footnote{Notice this would not happen with compactly supported detectors.}. We will have to precisely quantify these spurious signalling effects to make sure that when we talk about the entanglement acquired by the detectors we have genuine entanglement harvesting from the field.


Indeed, as we mentioned in Subsection \ref{sub:harvest}, it is possible to classify the entanglement acquired by the detectors into generated through signalling and harvested from the field state itself.  As discussed above, if $\mathcal{M}$ is dominated by the imaginary part contribution to the Wightman, the entanglement acquired is not harvested from the field, but rather due to communication~\cite{ericksonNew}.  In the setup of Fig. \ref{scheme}, when the interaction of the detectors with the field happens simultaneously, this classification can be done in terms of the real and imaginary parts of the $\mathcal{M}$ term of Eq. \eqref{eq:Mk}. The imaginary part of $\mathcal{M}$ is associated with communication (associated with $\langle[\hat{\phi}(\mf x),\hat{\phi}(\mf x')]\rangle$) and the real part of $\mathcal{M}$ is associated with the field's correlations (associated with $\langle\{\hat{\phi}(\mf x),\hat{\phi}(\mf x')\}\rangle$). We refer the reader to \cite{ericksonNew} for more details about the interpretation of these terms. In this setup, the real and imaginary parts of $\mathcal{M}$ can be written as
\begin{align} \label{eq:ReM}
    \Re\mathcal{M} &= -\frac{\lambda^2}{2 \pi^2}\int\frac{\dd \kappa}{2\omega_{\kappa}} \, \kappa^2 \,e^{-\kappa^2 \d^2}\, e^{-(\w^2+ \omega_{\kappa}^2)} \text{sinc}(\kappa \l) \nonumber, \\
    \Im\mathcal{M} &= \frac{\lambda^2}{2 \pi^2}\int\frac{\dd \kappa}{2\omega_{\kappa}} \, \kappa^2 \,e^{-\kappa^2 \d^2}\, e^{-(\w^2+ \omega_{\kappa}^2)} \text{sinc}(\kappa \l)  \\
    &\phantom{===================}\times \text{erf}\left( \ii  \omega_{\kappa} \right)\nonumber.
\end{align}

As stated above, we are interested in regimes in which the signalling between the detectors is negligible, so that all the entanglement is extracted from the field. In order to identify these regimes, we use the estimator of genuine harvesting from~\cite{ericksonNew}:
\begin{align} \label{eq:estimator}
    \mathcal{N}^{+} &= |\Re\mathcal{M}| - \mathcal{L}\\
    &= \!\frac{\lambda^2}{2 \pi^2} e^{-\w^2\!-\m^2 }\!\!\!\!\int\!\frac{\dd \kappa}{2\omega_{\kappa}} \, \kappa^2e^{-\kappa^2 }\!\!e^{-\kappa^2 \d^2}
    \!\!\left(\text{sinc}(\kappa \l)\! - \!e^{-2\w \omega_{\kappa}} \right). \nonumber
\end{align}
In the equation above, we used that the real part of $\mathcal{M}$ is negative, which gives $|\Re\mathcal{M}| = -\Re\mathcal{M}$. This allows to cast the negativity estimator as a single momentum integral.

\begin{figure}[h] 
    \includegraphics[width=8.7cm]{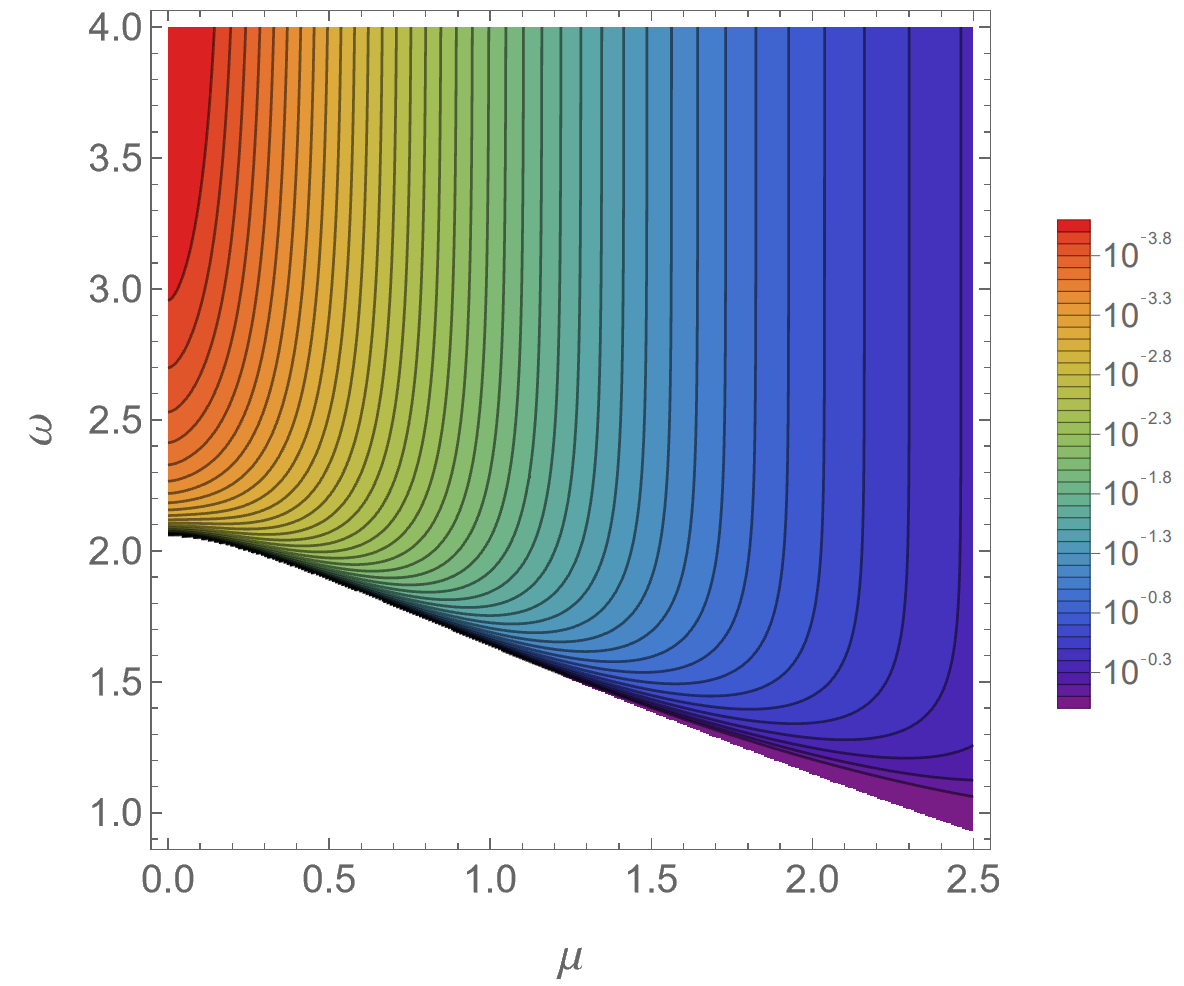}
    \caption{Relative error between negativity and the approximation in Eq. \eqref{eq:ReM}, in which the imaginary part of $\mathcal{M}$ is neglected. The separation between the detectors was $\l=5$ and the detectors' size $\sigma = 0.2$. The white region corresponds to values of the parameters where negativity is identically zero.}\label{fig:errorNeg2}
\end{figure}

In Fig. \ref{fig:errorNeg2} we show the relative difference between the second order full negativity $\mathcal{N}$ and the estimator $\mathcal{N}^+$, which neglects the signalling contribution, for $\l = 5$ and $\sigma = 0.2$. We find that for small values of the field mass, the entanglement acquired by the particle detectors corresponds to genuine entanglement harvesting. As the mass of the field increases and for small values of the energy gap, communication starts contributing to the entanglement harvested by the detectors. In this paper we will mostly consider values for $\m$ smaller than $2$, for which it is safe to assume that the entanglement acquired by the detectors is mostly genuinely harvested.


Overall, we conclude that for small detector gaps the choices of spacetime smearing functions of Eqs. \eqref{lambdaA} and \eqref{lambdaB} allow the detectors to signal even if they are in spacelike separated regions. Nevertheless, if $\omega$ is large enough, the signalling contribution is negligible compared to the true entanglement harvested from the field's state.


\section{The effect of mass on Entanglement Harvesting}\label{mass}

In this section we study in detail the effects of the field's mass on the protocol of entanglement harvesting. For concreteness, we will focus on the Gaussian switching and smearing functions\footnote{Gaussian switching are good representatives for smooth detector smearing. One would not expect any qualitative differences given by the shape of the detectors as long as we consider smooth smearing and switching~\cite{mckay}} outlined in Section \ref{protocol}. In Subsection \ref{sub:mas} we study how the field's mass influences the negativity acquired by a given pair of detectors. In Subsection \ref{sub:optimizingMass} we consider detectors which optimize the harvested negativity, and conclude that entanglement harvesting can be enhanced by considering fields with small mass. 


\subsection{The effect of mass in entanglement harvesting.}\label{sub:mas}

\begin{figure}[h]
    \includegraphics[scale=0.73]{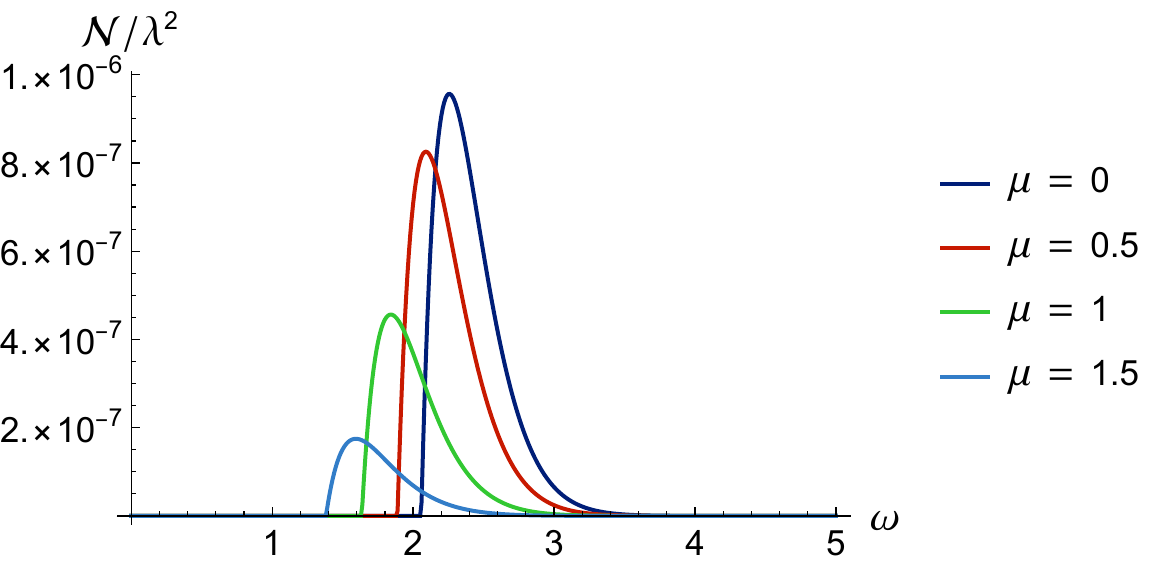}
    \caption{Negativity as a function of the energy gap of the detectors for different values of the mass of the field. The distance between the detectors was chosen as $\l = 5$, and the detectors' size $\d = 0.2$. }\label{Omega}
\end{figure}

\begin{figure}[h]
    \includegraphics[scale=0.73]{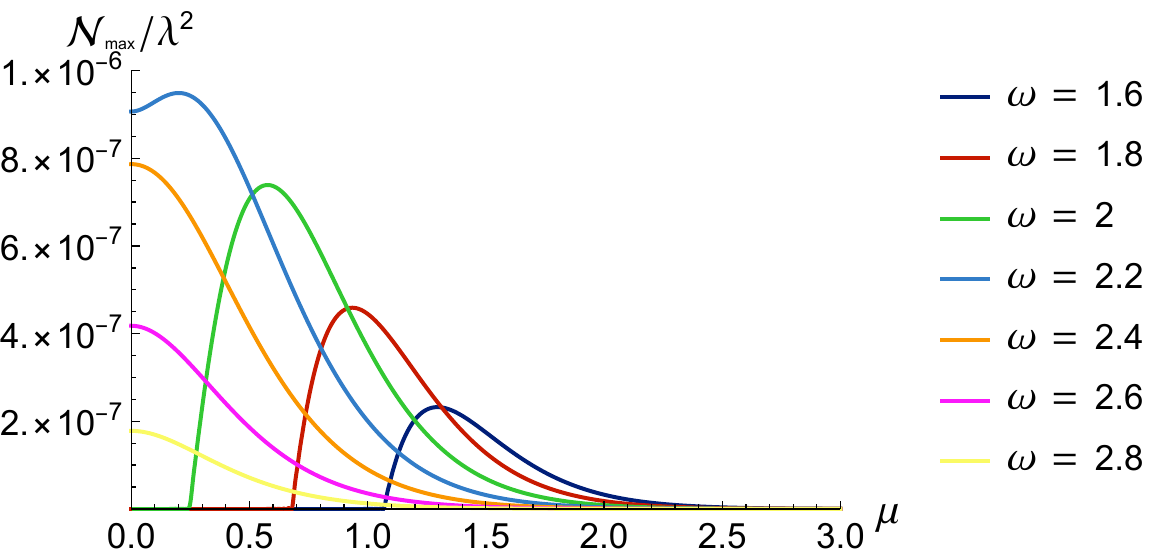}
    \caption{Negativity as a function of the field mass for different values of the energy gap. The distance between the detectors was fixed as $\l = 5$ and the detectors size $\d = 0.2$.}\label{m}
\end{figure}

\begin{figure}[h!]
    \includegraphics[width=8.5cm]{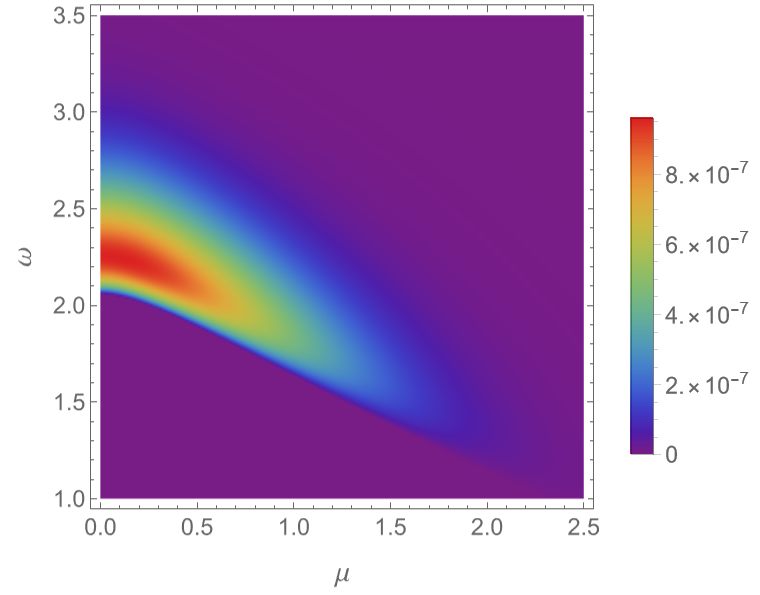}
    \caption{Negativity as a function of the energy gap of the detectors and the mass of the field. The distance between the detectors was fixed as $\l = 5$ and the detectors size $\d = 0.2$. Notice that this plot also includes regions where entanglement is affected by communication (See Fig. \ref{fig:errorNeg2}).
    }\label{2D}
\end{figure}

 
 In this Subsection we analyze the behaviour of the entanglement acquired by the detectors as a function of the parameters $\w = \Omega T$ and $\m = m T$. In Figs. \ref{Omega}, \ref{m} and \ref{2D} we plot the negativity of the detectors for varying values of $\mu$ and $\w$ when the detectors are separated by a distance $\l = 5$ and have size $\d = 0.2$, so that the interaction regions can be effectively considered spacelike separated and we can use the analysis of Section~\ref{protocol}.

In Fig. \ref{Omega} we plot the entanglement extracted by the detectors as a function of $\w$ for different values of mass. As is usually seen in the literature, we see that for small enough values of $\w$ it is not possible to harvest entanglement from the field, until a finite threshold $\w$ is reached. After this threshold, the negativity peaks and decreases monotonically. In these plots we see that as the field's mass increases, the maximum amount of negativity that can be harvested decreases. On the other hand, we see that increasing field masses allow one to harvest entanglement using detectors with smaller gaps.

In Fig. \ref{m} we plot the detectors negativity as a function of the field's mass for varying values of $\w$. We can clearly see two regimes with different behaviours\footnote{We note that in Subsection \ref{sec:optimizing} we will be able to precisely quantify the value of $\w$ that controls this change of behaviour in terms of $\l$, and $\d$.}: $\w\lesssim 2.3$ and $\w\gtrsim 2.3$. When $\w \lesssim 2.3$, we find a regime where the negativity increases with mass. Moreover, if $\w$ is small enough, we see regimes where it is impossible to harvest from fields with small mass, but as the mass increases, it becomes possible to harvest entanglement. This result is perhaps surprising since we know that the field correlations decay exponentially fast with mass, and yet, we see regimes where the field's mass can in fact increase the amount of entanglement that can be harvested from the field. A similar effect was also seen in \cite{carol}, using detectors in different initial states. When $\w \gtrsim 2.3$ we see a monotonic decay of the negativity with the field's mass. Indeed as the mass of the field goes above all the other scales in the problem the ability of the detectors to harvest entanglement is lost. As we will argue in Sec.~\ref{sec:optimizing}, the change of behaviour we observed in these plots happens when $\w \approx \l/2$.

\begin{figure}[h]
    \includegraphics[scale=0.73]{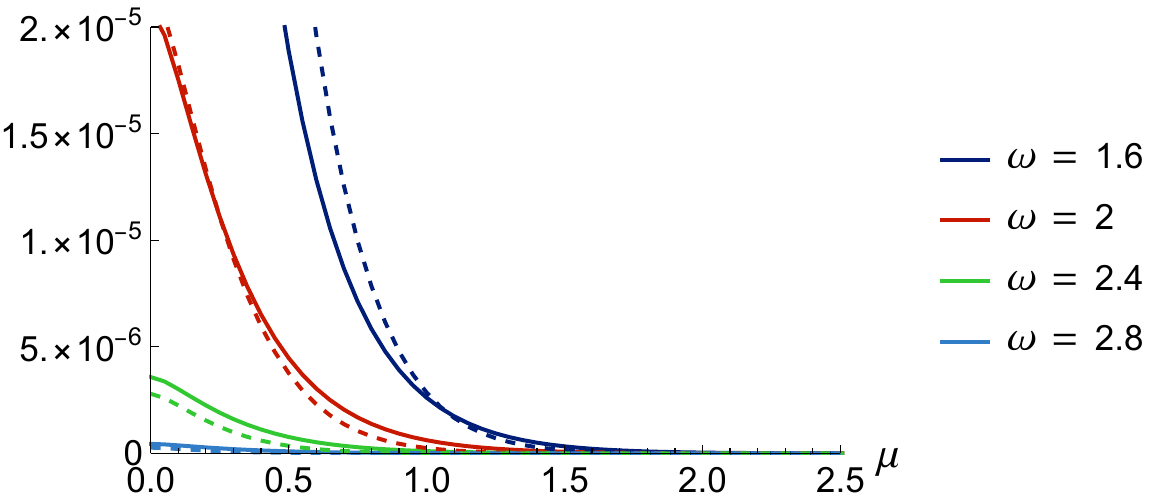}
    \caption{$|\mathcal{M}|$ and $\mathcal{L}$ as a function of the field's mass for different values of $\w$. The dashed lines correspond to $\mathcal{L}$ and the solid lines to $|\mathcal{M}|$. The distance between the detectors was fixed as $\l = 5$ and the detectors size as $\d = 0.2$. 
    } \label{ML}
\end{figure}
Since the correlations in the field decrease with mass, the only possible explanation for a non-monotonic behaviour of harvesting with mass is that the local noise $\mathcal{L}$ of the detectors  is suppressed  faster than the correlation terms $\mathcal{M}$ as the mass increases. In order to better understand this behaviour, we plot $|\mathcal{M}|$ and $\mathcal{L}$ as a function of $\mu$ for different values of $\w$ in Fig. \ref{ML}. The dashed lines correspond to values of $\mathcal{L}$ and the solid lines to values of $|\mathcal{M}|$. We can then see that for $\w < 2.3$, the $\mathcal{L}$ terms starts larger, but decreases faster, so that eventually the $\mathcal{M}$ term catches up. This results in the peaks we observed in Fig \ref{m}. For $\w > 2.3$, we see that for every mass, we have $|\mathcal{M}|>\mathcal{L}$, and the negativity behaves monotonically with mass since $|\mathcal{M}|$ decays faster than $\mathcal{L}$.

To have a better impression of the big picture, in Fig. \ref{2D}, we also plot the negativity of the two-detectors system as a function of both the field's mass and the detectors gap setting $\l = 5$ and $\sigma = 0.2$ covering all the regimes analyzed above. 

\subsection{Optimal detectors and the effect of mass}\label{sub:optimizingMass}

We have seen that for particle detectors with a fixed gap $\w$, there is a particular value of the mass of the field (often non-zero) that maximizes the extraction of entanglement. One could wonder whether this is because in these cases the gap of the detector is poorly chosen. Thus, one might expect that by choosing the optimal value of the gap so that entanglement harvesting is maximized for each $\mu$, the dependence of entanglement on the field mass will be monotonically decreasing, tracking the mass dependence of the field correlations. In this Subsection we will show that this is not the case.

\begin{figure}[h!]
    \includegraphics[scale=0.73]{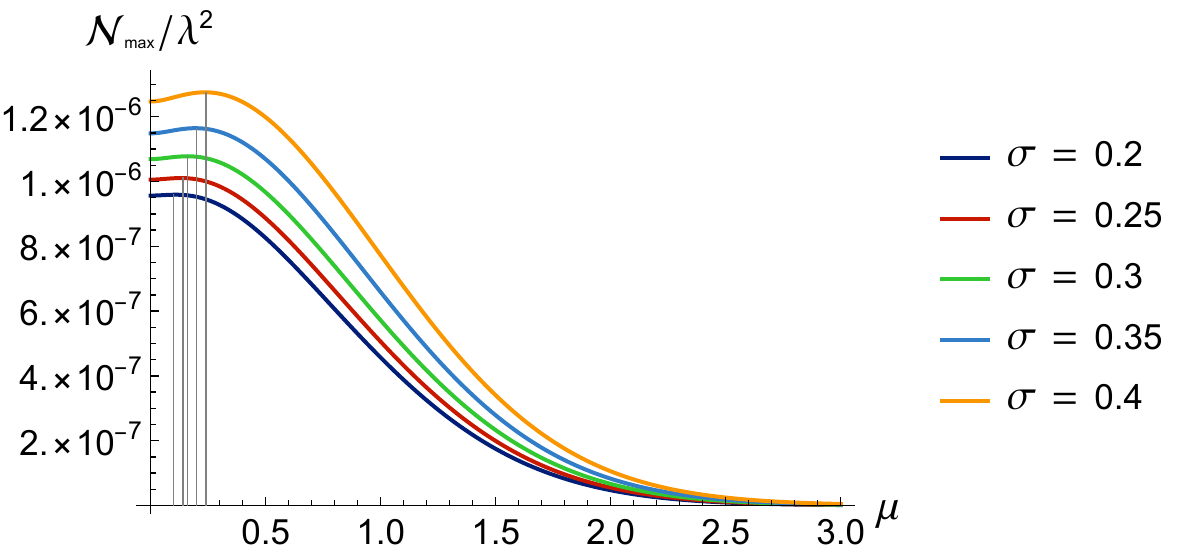}
    \caption{Negativity for the maximizing value of $\w$ as a function of the field's mass for varying values of $\d$ which still ensure mostly spacelike separation. We fixed $\l = 5$ for these plots. The vertical lines correspond to the maximum value of each curve.}\label{Nmaxmasssigma}
\end{figure}
\begin{figure}[h!]
    \includegraphics[scale=0.73]{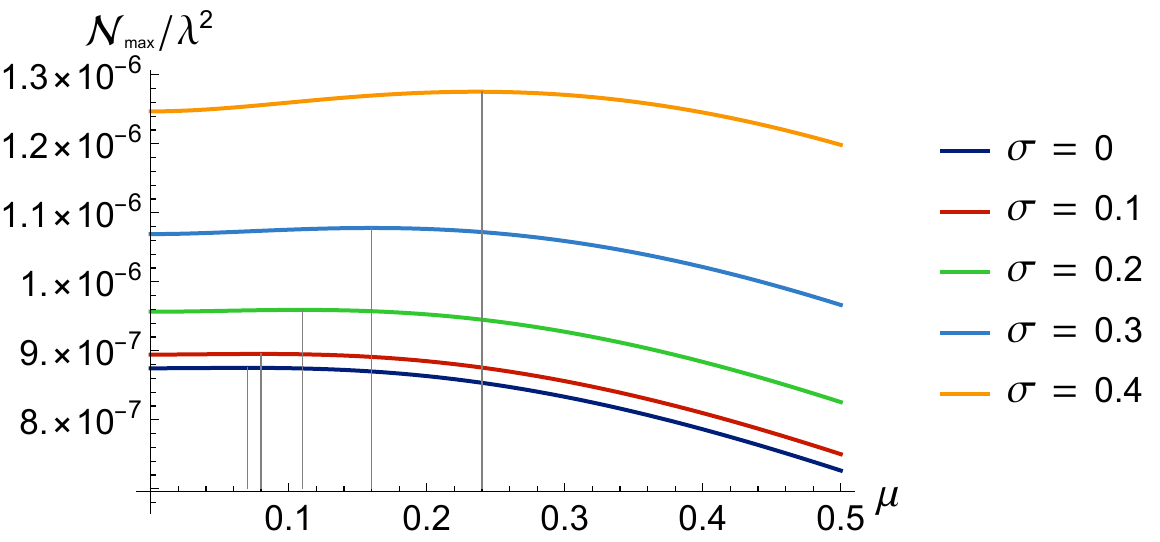}
    \caption{Negativity for the maximizing value of $\w$ as a function of the field's mass for varying values of $\d$, starting at with a pointlike detector. We fixed $\l = 5$ for these plots and focused on the small mass behaviour, close to the peaks of the plot. The vertical lines correspond to the maximum value of each curve.}\label{zoom}
\end{figure}
\begin{figure}[h]
    \includegraphics[scale=0.73]{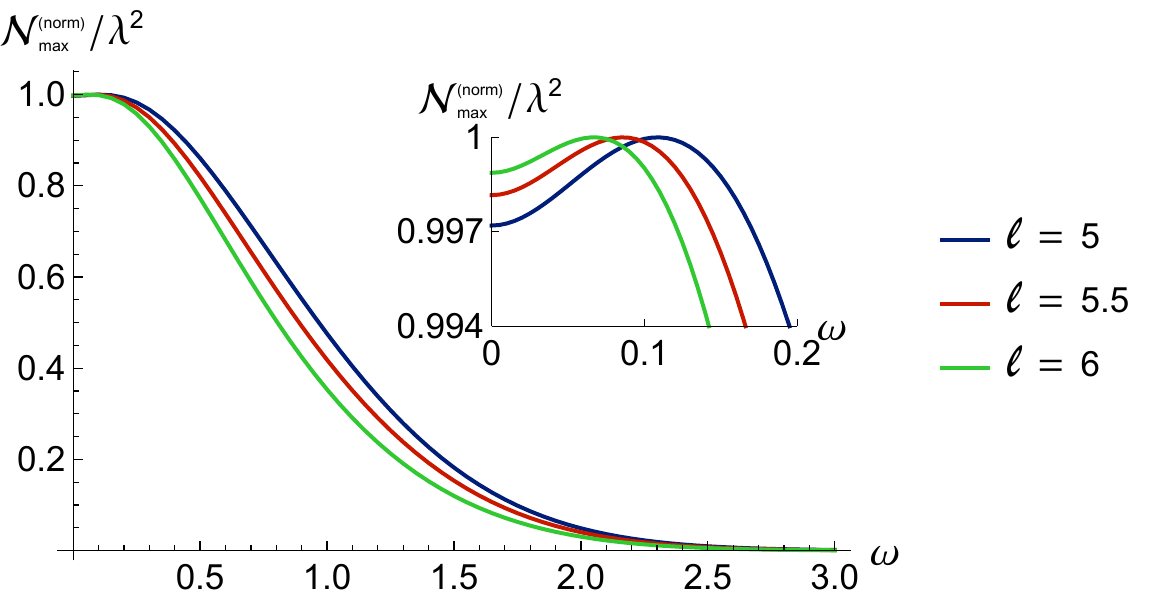}
    \caption{Negativity (normalized by its peak value) for the maximizing value of $\w$ as a function of the field's mass for varying values of $\l$ which still ensure mostly spacelike separation. We fixed $\d= 0.2$ for these plots.  
    \label{NmaxmassL} 
    }
\end{figure}
\begin{figure}[h]
    \includegraphics[scale=0.73]{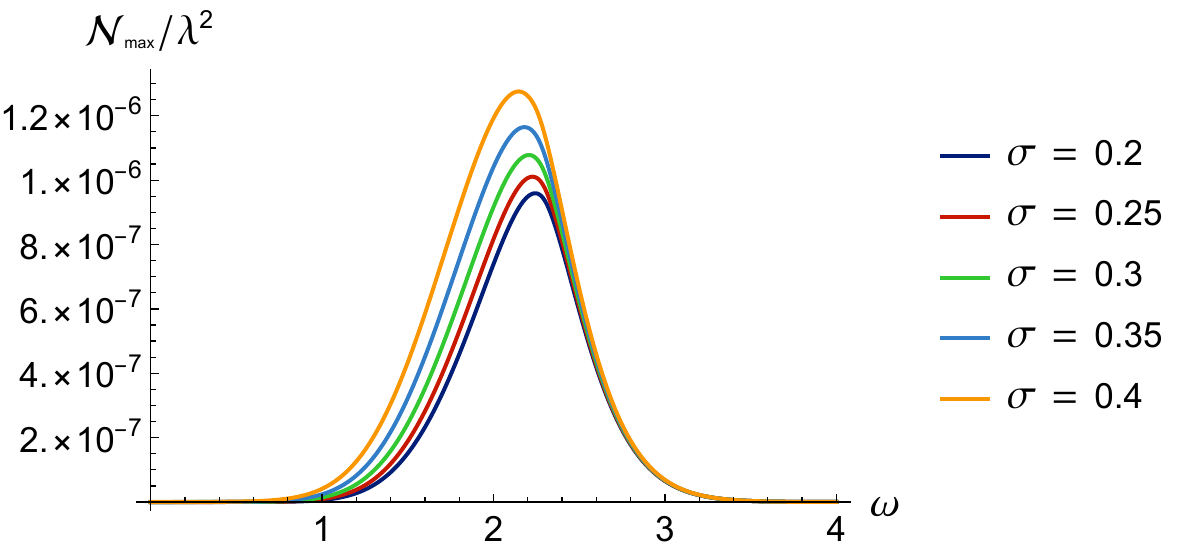}
    \caption{Negativity for the maximizing value of mass as a function of the detectors' gap for varying values of $\d$ which still ensure mostly spacelike separation. We fixed $\l = 5$ for these plots. }\label{NmaxOmegasigma}
\end{figure}
\begin{figure}[h!]
    \includegraphics[scale=0.73]{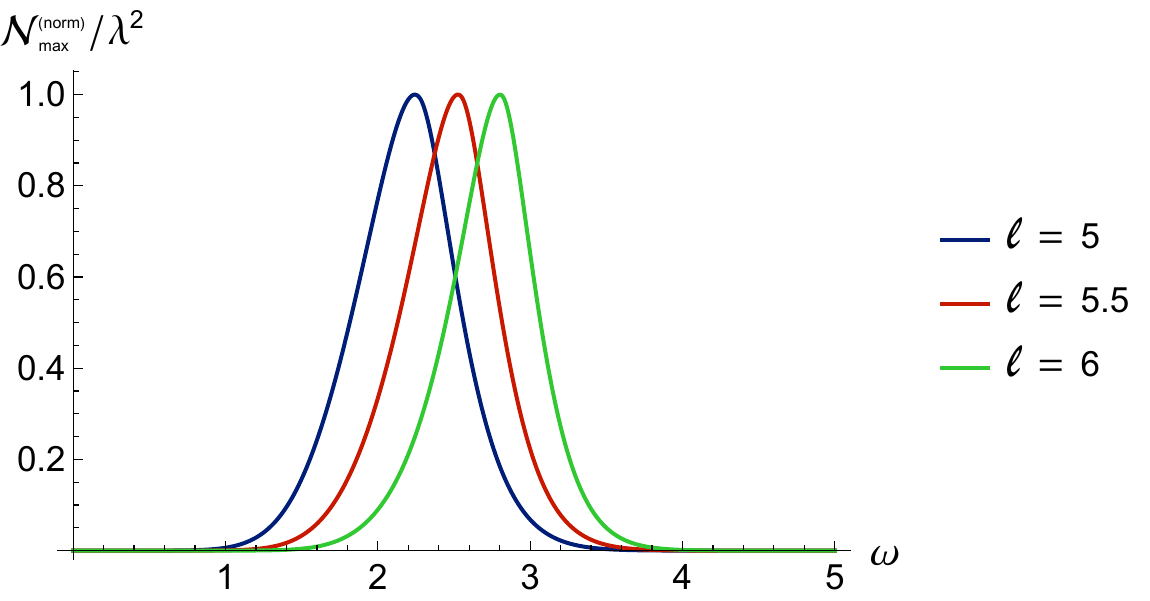}
    \caption{Negativity (normalized by its peak value) for the maximizing value of mass as a function of the detectors' gap for varying values of $\l$ which still ensure mostly spacelike separation. We fixed $\d = 0.2$ for these plots.}\label{NmaxOmegaL}
\end{figure}

In order to study the maximum amount of entanglement that can be extracted by \emph{any} two Gaussian-smeared (effectively) spacelike separated detectors, we plot the optimal negativity (by setting the detectors' gap to the value that maximizes $\mathcal{N}$ for each mass) as a function of $\mu$ in Figs. \ref{Nmaxmasssigma}, \ref{zoom} and \ref{NmaxmassL}. Additionally we study the optimization of harvesting with respect to the field mass (keeping the other parameters constant for the optimization) as a function of the detectors' gap in Figs. \ref{NmaxOmegasigma} and \ref{NmaxOmegaL}.

In Fig. \ref{Nmaxmasssigma} (and the magnified version in Fig.~\ref{zoom}) we see the behaviour of the negativity maximized over $\w$ for different values of the detector size and a fixed detector separation of $\l = 5$. Overall, we see that even after choosing the optimal $\w$ that maximizes negativity, a small non-zero mass yields more entanglement than the massless case. The peaks on the plot are larger, and shifted towards larger values of mass for larger values of $\sigma$. Although in Fig. \ref{Nmaxmasssigma} it may seem that the peaks might disappear as the detector size goes to zero, this is not the case. In Fig. \ref{zoom} we display the case of pointlike detectors with $\sigma = 0$. We still find a  maximum for non-zero mass even in this case. 

In Fig. \ref{NmaxmassL} we plot the negativity (normalized to its peak value\footnote{The negativity decays exponentially with $\ell$, as already seen in \cite{Pozas-Kerstjens:2015}. Since we want to see for what value of $\mu$ the negativity peaks it is convenient to normalize the negativity as function of $\mu$ by its maximum value.}) for the value of $\w$ that maximizes it as a function of the field mass and for different detector separations $\l$, with fixed detector size $\d = 0.2$. In this plot we observe a similar behaviour to that of Figs. \ref{Nmaxmasssigma} and \ref{zoom}, where the negativity peaks for small nonzero values of the field's mass. In summary, we find that $\l$ does not change the overall behaviour of the maximized negativity as a function of mass.

In Fig. \ref{NmaxOmegasigma} we plot the negativity for the value of the  the field's mass that maximizes the entanglement harvested as a function of $\w$. We fix $\l = 5$ and consider different values of $\d$ which are small enough to ensure that the interaction regions are spacelike separated. We see resonance-like behaviour. As we will discuss later, the peaks in the negativity happen at approximately $\w  \approx \l/2+\mathcal{E}(\ell)+\mathcal{A}(\ell)\d^2+\mathcal{B}(\ell)\m^2$ for a negative function $\mathcal{A}(\ell)\d^2$ (see Appendix \ref{app:heitor} for details).

In Fig. \ref{NmaxOmegaL} we also plot the negativity as a function of $\w$ when the mass is chosen to maximize the negativity for varying values of $\l$. We picked \mbox{$\d = 0.2 $}, ensuring approximate spacelike separation. We also see the resonance behaviour, with peaks of negativity for \mbox{$\w  \approx \l/2+\mathcal{E}(\ell)+\mathcal{A}(\ell)\d^2+\mathcal{B}(\ell)\m^2$} for a negative function $\mathcal{A}(\ell)\d^2$ (see Appendix \ref{app:heitor} for details).

\begin{figure}[h!]
    \includegraphics[scale=0.73]{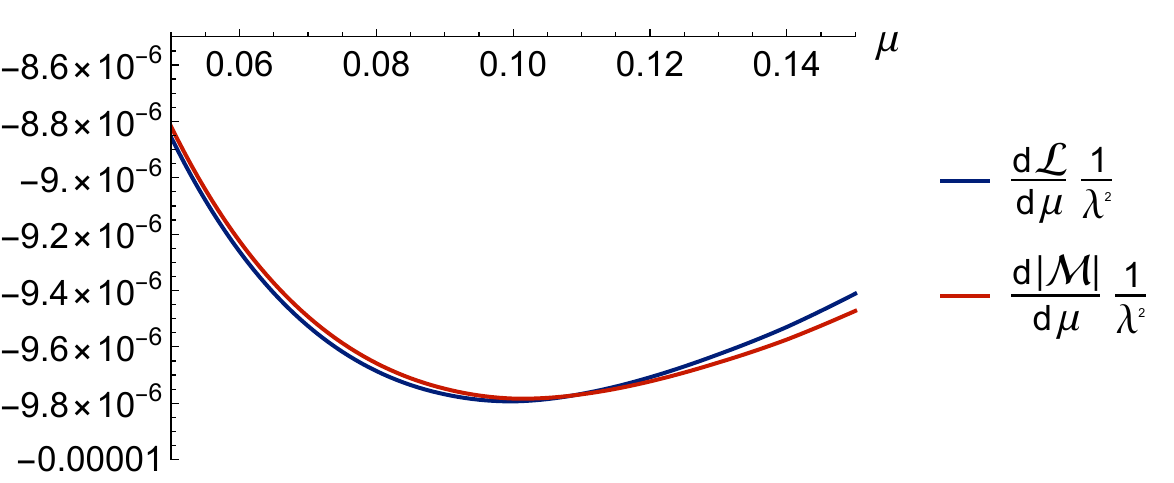}
    \caption{Derivatives of $\mathcal{L}(\w_{max}(\m))$ and $|\mathcal{M}(\w_{max}(\m))|$  as a function of the adimensional mass $\mu$. We fixed $\l=5$ and $\d = 0.2$ for these plots. 
    }
\label{lines2?}
\end{figure}

Finally, in Fig. \ref{lines2?} we show the derivatives of $\mathcal{L}$ and $|\mathcal{M}|$ with respect to the mass of the field always choosing the value of the gap that maximizes negativity for that mass, $\w_{\text{max}}(\m)$. The point where the derivatives of  $|\mathcal{M}|$ and $\mathcal{L}$ cross corresponds to the peak of negativity.  This showcases that as mass increases, the noise and correlation terms are affected differently. For small mass the noise decays faster than the correlation term as the mass increases, leading to a maximum of harvested negativity for some finite value of mass. 

We conclude that in the small mass regime it is possible to harvest more entanglement from a massive field than from a massless field, even though the field itself contains less entanglement between the interaction regions. While the theoretical entanglement of the field between the two spacelike separated regions decreases with mass, the entanglement that can be accessed by physical systems does not share the same monotonical behaviour. 

\section{Optimizing parameters for entanglement harvesting}\label{sec:optimizing}

In this section we analyze the fact that that $\w \approx \l/2$ maximizes the harvested entanglement. In fact, we will be able to find an approximate expression for the detector gap that maximizes negativity as a function of its size and the field mass. The discussion in this section will be focused in regimes of small mass and $\w \geq  2$ so that we can safely replace the $\mathcal{M}$ term with its real part. This means that under this approximation Eq. \eqref{eq:estimator} can be taken as the negativity. 

We start by analyzing the massless case, where  $\omega_{\kappa} = \kappa$, and the integrals for $\Re \mathcal{M}$ and $\mathcal{L}$ can be solved analytically, yielding
\begin{equation}
    \begin{gathered}
    |\Re\mathcal{M}|_{\mu=0} = \frac{\lambda^2}{8 \pi^{3/2}\l} e^{-\w-\frac{l^2}{4}}  \text{erfi}\left(\frac{\l}{2}\right).\\
    \mathcal{L}_{\m = 0} = \frac{\lambda^2e^{-\w^2 }}{8\pi^2}\left(1-\sqrt{\pi}\w e^{\w^2 }\text{erfc}(\w)\right).
    \end{gathered}
\end{equation} Then the harvested negativity in Eq.~\eqref{eq:estimator} can be expressed as a function of $\w$ and $\l$:
\begin{align}
    \mathcal{N}^{+}_{\m=0} \!\!\approx \!\frac{\lambda^2e^{-\w^2}}{8\pi^2} \!\bigg(\!\frac{\sqrt{\pi} }{\l}e^{-\frac{\l^2}{4}}\label{m0Hector} \!\text{erfi}\left(\l/2\right)\!+\!\sqrt{\pi}\w e^{\w^2 }\!\text{erfc}(\w)\!- \!1 \!\bigg).
\end{align}
In order to see what value of the gap $\w$ yields a maximum for the negativity, we can differentiate $\mathcal{N}_{\mu=0}^{+}$ and look for its zeros. We find:
\begin{equation}
    \pdv{\mathcal{N}^{+}_{\,\m=0}}{\w} = \frac{\w}{8\pi^{3/2}}\!\!\left(\!\frac{\text{erfc}(\w)}{\w} -\frac{2 }{\l}e^{-\frac{\l^2}{4} - \w^2}\!\text{erfi}\left(\frac{\l}{2}\right)\!\!\right)\!.
\end{equation}
Setting the expression above to zero yields the following relationship between $\w $ and $\l $:
\begin{equation}
    \frac{e^{\w^2} \text{erfc}(\w)}{\w} = \frac{e^{-\left(\frac{\l}{2 }\right)^2}\text{erfi}(\l/2)}{\l/2}.
\end{equation}
{It is possible to show\footnote{The asymptotic expansions of $f(u)$ and $g(u)$ read
\begin{equation*}
    f(u) \sim \frac{1}{\sqrt{\pi} u^2} \sum_{k=0}^\infty (-1)^k\frac{(2k-1)!!}{(2 u^2)^{k}},\:\:  g(u) \sim \frac{1}{\sqrt{\pi} u^2} \sum_{k=0}^\infty \frac{(2k-1)!!}{(2 u^2)^{k}},
\end{equation*}
so that $f(u) = g(u) + \mathcal{O}(u^{-4})$.} that the functions $f(u) =  \text{erfc}(u)e^{u^2}/u$ and $g(u) = \text{erfi}(u)e^{-u^2}/u$ behave similarly for large values of $u$. In particular, this implies that for large enough values of $\l$, the peaks of negativity happen for $\w \approx \l/2$, in agreement with our previous discussions.}

\begin{figure}[h]
    \includegraphics[scale=0.73]{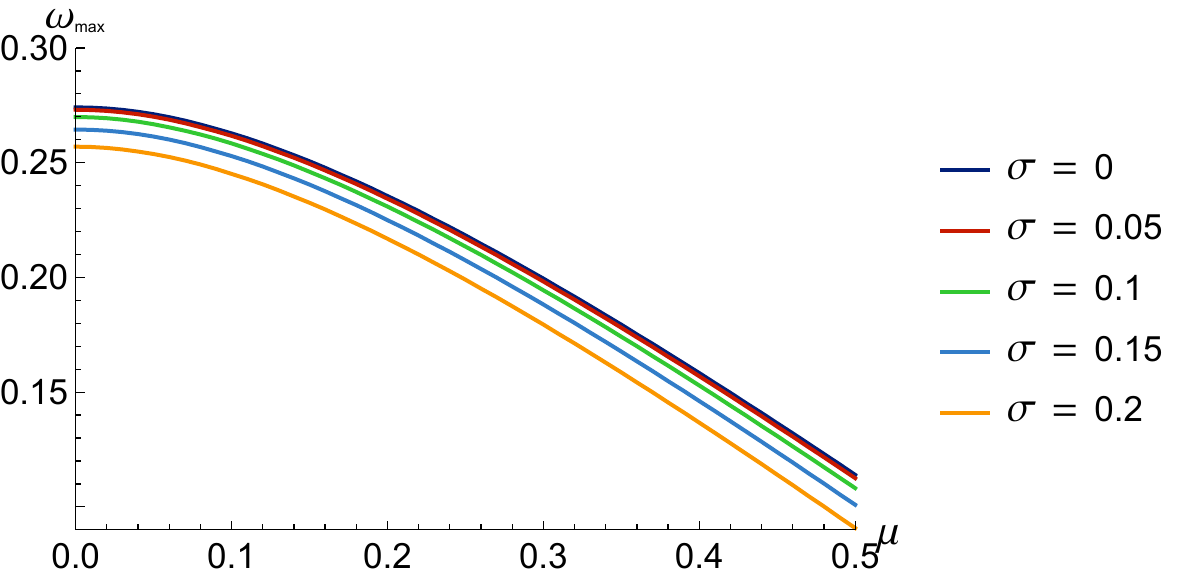}
    \caption{Energy gap $\w$ that maximizes the negativity as a function of the mass $\mu$ for different detector sizes.}\label{Hector<3}
\end{figure}

Moreover, in Appendix \ref{app:heitor}, we find that for fixed $\l$ and sufficiently small $\m$ and $\d$, the values of the detectors gap that maximize entanglement harvesting approximately satisfy
\begin{equation}\label{claim}
    \w_{\text{max}} \!\approx \frac{\l}{2} +\mathcal{E}(\ell)   + \mathcal{A}(\l)\d^2+\mathcal{B}(\l)\m^2,
\end{equation}
where approximate expressions for $\mathcal{E}(\ell)$, $\mathcal{A}(\l)$ and $\mathcal{B}(\l)$ can be found in Appendix \ref{app:heitor}. 
The result of Eq. \eqref{claim} is consistent with the behaviour found in Figs. \ref{NmaxOmegasigma} and \ref{NmaxOmegaL} if $\m$ and $\d$ are small enough. In fact, we have seen in Figs. \ref{Nmaxmasssigma}, \ref{zoom} and \ref{NmaxmassL} that the maximum of negativity as a function of mass happens for small masses ($\m \leq 0.25$). In Fig. \ref{Hector<3} we see how  the exact value of $\w$ that maximizes negativity  behaves as a function of the field mass. We see the decaying behaviour with mass expected from the approximation in Eq. \eqref{claim}.

For small field mass, Eq. \eqref{claim} is also helpful to explain the change from increasing to decreasing negativity as a function $\mu$ seen in Fig. \ref{m}. In Appendix \ref{app:heitor}, we find that the harvested negativity admits a power expansion in the field mass of the form \begin{equation}
    \mathcal{N}^{+} = \mathcal{N}^{+}_{\m = 0} + \frac{1}{2}\m^2 \left.\pdv{^2\mathcal{N}^{+}}{\mu^2}\right|_{\mu=0} + \mathcal{O}(\mu^4).
\end{equation} 
By analyzing the dependence of $\partial^2\mathcal{N}^{+}\!/\partial\mu^2|_{ \mu =0}$ on $\w$, $\l$ and $\d$, we conclude that this term changes sign at approximately $\omega_{\text{max}}(\mu = 0)$ from Eq. \eqref{claim}. With this we are able to approximately quantify the change in behaviour seen in Fig. \ref{m}, when the detector gap crosses $\l/2 + \mathcal{E}(\l)$, which for $\l = 5$ yields $\w\approx 2.3$.

Overall, Eq. \eqref{claim} yields a good approximation for the gap of the detectors that will maximize entanglement harvesting as a function of their size, separation and field mass.



\section{Conclusions}\label{conclusions}

We have performed a detailed study of the effect of the field mass and the effect of the detector gap on the protocol of entanglement harvesting. We found several results that challenge previous intuition on how entanglement harvesting should behave as a function of the field mass. Namely, even though the field correlations decrease as the mass of the field increases, there are regimes where the field's mass can increase the amount of entanglement harvested by two spacelike separated detectors. Furthermore we have found that for fields of small mass (including massless fields) the amount of entanglement harvested can be optimized by choosing the detector gaps to match the scale of the spatial separation between the detectors.  


In particular, we found that the field mass can enhance entanglement harvesting in two cases. First, if one fixes the detectors' gap, there is an optimal non-zero mass of the field that maximizes the amount of entanglement that can be harvested. What is more, one can find regimes where detectors with a fixed gap cannot harvest entanglement for massless fields but a finite field mass allows for entanglement extraction. Second, even when one chooses the optimal value of the detectors' gap that maximizes entanglement for each field mass, we found that more entanglement can be harvested from a field with a small mass than from a massless field. Considering that entanglement harvesting is a competition between the local noise that detectors experience and the non-local field correlations, we traced back the increase of entanglement with mass to the fact that (for small masses) the local noise terms are suppressed with the field mass in a stronger manner than the field correlations.

In summary, while it is well known that the correlations of a quantum field decrease with its mass (and so does the entanglement between spacelike separated regions), we showed  that if one attempts to extract entanglement from a quantum field, small masses can actually improve the protocol of entanglement harvesting by decreasing the noise experienced by the probes. That is, although the field itself contains less entanglement, the physical systems that can be used to extract these quantum correlations can benefit from a small field's mass for extracting entanglement.


\section*{Acknowledgements}

The authors thank Achim Kempf for kindly providing space at the physics of information laboratory where
part of this research was conducted. H. M.-G. has been
funded by the mobility grants program of Centre de
Formació Interdisciplin\`aria Superior (CFIS) - Universitat
Polit\`ecnica de Catalunya (UPC). T. R. P. acknowledges
support from the Natural Sciences and Engineering
Research Council of Canada (NSERC) via the Vanier
Canada Graduate Scholarship. E. M.-M. is funded by the
NSERC Discovery program as well as his Ontario Early
Researcher Award. Research at Perimeter Institute is
supported in part by the Government of Canada through
the Department of Innovation, Science and Industry
Canada and by the Province of Ontario through the
Ministry of Colleges and Universities.

\appendix
\section{Entanglement harvesting with different detector gaps}\label{app:Omegas}

In this Appendix we consider the protocol of entanglement harvesting from the Minkowski vacuum using two inertial comoving particle detectors with different gaps $\Omega_{\textsc{a}}$ and $\Omega_{\textsc{b}}$. We assume the spacetime regions of interaction to be given by Gaussians according to the protocol outlined in Section \ref{protocol}, so that the spacetime smearing functions of the detectors can be written according to Eqs. \eqref{lambdaA}, \eqref{lambdaB}, \eqref{chi} and \eqref{F}. 

We consider the protocol of entanglement harvesting when the spacetime smearing of the interaction of each detector is approximately spacelike separated, with $\l = 5$. This ensures that the entanglement acquired by the detectors is overwhelmingly due to the correlations previously present in the quantum field. In order to screen out any effects related to the variation of the total energy of the system, we  parametrize the gaps in a way which keeps the sum of the detectors gap constant, while only varying the difference between the gaps. This can be accomplished by defining $\overline{\w} = \frac{1}{2}(\w_{\textsc{a}} + \w_{\textsc{b}})$ as the average detector gap and $\delta \w = \frac{1}{2}(\w_{\textsc{a}} - \w_{\textsc{b}})$, so that $\w_{\textsc{a}} = \overline{\w} + \delta \w$ and $\w_{\textsc{b}} = \overline{\w} - \delta \w$. 
In order to study the difference in the detector gaps, one would keep $\w$ constant and vary $\delta \w$.

\begin{figure}[h!]
    \includegraphics[scale=0.73]{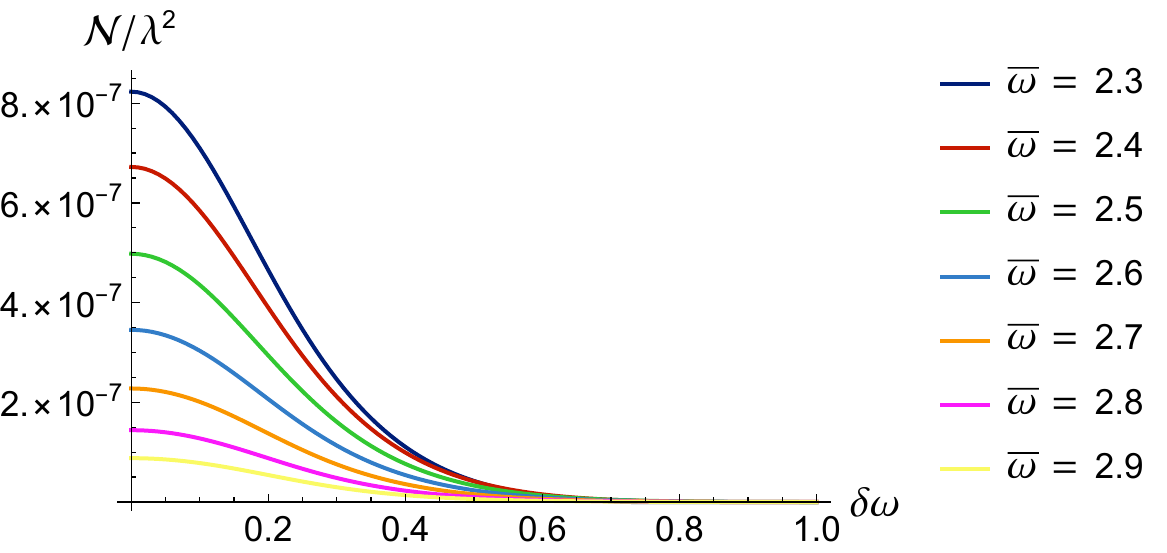}
    \caption{Negativity extracted by the detectors as a function of the gap difference $\delta \w$. We fixed the detector separation as $\l = 5$ and the the detector size $\sigma = 0.2$ for a massless field ($\mu = 0$).}\label{fig:diffOmega}
\end{figure}

With these conventions, it it possible to write the non-local term $ \mathcal{M}$ as
\begin{align} 
    \mathcal{M} &= \frac{\lambda^2}{4 \pi^2}\int_{0}^{\infty}\frac{\dd \kappa}{2\omega_{\kappa}} \,\kappa^2 \,e^{-\kappa^2 \d^2}\,  \text{sinc}(\kappa \l) \\
    &\times\! e^{-(\overline{\w}^2 +\omega_{\kappa}^2 +\delta\w^2)}\!\big(e^{-2\omega_{\kappa} \,\delta \w} \left(1\!- \!\ii \, \text{erfi}\left( (\omega_{\kappa} \!+\! \delta \w)\right) \right)  \nonumber \\ &\phantom{\!=e^{-(\w^2 +\omega_{\kappa}^2 +\delta\w^2)}}+e^{2\omega_{\kappa} \,\delta \w}\!\! \left(1\!-\! \ii \, \text{erfi}\left( (\omega_{\kappa} \!- \!\delta \w) \right) \right) \!\big) .\nonumber
\end{align}
Notice that the expression above is even with respect to $\delta \w$, which is natural since no detector should be privileged. Now, taking into account that the detectors are not identical, the negativity of the two-detector system cannot be simplified to yield $\text{max}(0,|\mathcal{M}| - \mathcal{L})$. Instead, we must use the full expression from Eq. \eqref{eq:negativeEigenvalue}. With these expressions, we can plot the behaviour of the negativity of the two-detector system as a function of the gap difference. In Fig. \ref{fig:diffOmega} we plot the negativity as a function of $\delta \w$ for multiple values of $\overline{\w}$, while considering a massless scalar field and both detectors of size of $\d = 0.2 $ separated by a distance $\l = 5$. It is possible to see that the negativity is a monotonically decreasing function of the detectors gap difference, $\delta\w$. In other words, one can say that a resonance effect happens when $\w_{\textsc{a}} = \w_{\textsc{b}}$, which maximizes entanglement harvesting. Overall, we find that considering comoving detectors with different energy gaps cannot increase entanglement harvesting. Notice that this is in direct contradiction with one of the main claims in~\cite{Hui}. The likely reason why they find a different result in this analysis is  because they do not consider the sum of the two detector gaps constant when they perform their study, and increasing the total gap is well known to enhance the amount of entanglement harvested from the field in some regimes (see, e.g.,~\cite{Pozas-Kerstjens:2015}).

\section{Second order expansion in the mass for the negativity}\label{app:heitor}

In this appendix, we compute the Taylor expansion of the negativity to second order in the mass of the field $\mu$ and to second order in the detector size $\sigma$. We use this expansion to justify the observed behaviour of negativity in Fig.~\ref{m} for small masses. In order to ease the calculations, we first analyze the case in which the detectors are pointlike $\sigma = 0$.

We work in the regimes identified in subsection \ref{spacelike} in which the entanglement acquired through signalling between the detectors is negligible compared to the entanglement harvested  from the field. The negativity is very well approximated in these regimes by Eq. \eqref{eq:estimator}. We proceed perturbatively in the dimensionless mass $\mu = mT$, so that we can write
\begin{widetext}
\begin{align}
    \mathcal{N}^{+} = \mathcal{N}^{+}_{\m = 0}+ \m\left.\frac{\partial \mathcal{N}^{+}}{\partial \m}\right|_{\m = 0}+ \frac{\m^2}{2}\left.\frac{\partial^2 \mathcal{N}^{+}}{\partial \m^2}\right|_{\m = 0} + \mathcal{O}(\mu^3).
\end{align}
The massless term can be obtained by direct integration of Eq. \eqref{eq:estimator}, which yields
\begin{align}
 \mathcal{N}^{+}_{\m=0} \approx \frac{\lambda^2e^{-\w^2}}{8\pi^2} \bigg(&\frac{\sqrt{\pi} }{\l}e^{-\frac{\l^2}{4}} \text{erfi}\left(\l/2\right)+\sqrt{\pi}\w e^{\w^2 }\text{erfc}(\w)- 1 \bigg).\label{eq:0Derivative}
\end{align}
In order to compute the derivatives of $\mathcal{N}^+$ with respect to the $\mu$, we differentiate Eq. \eqref{eq:estimator} under the integral sign. The first derivative yields 
\begin{equation} \label{eq:firstDerivative}
\frac{\partial \mathcal{N}^{+}}{\partial \m} = \lambda^2 \frac{\m}{4\pi^2} \int_{0}^{\infty}\frac{\dd \kappa}{\omega_{\kappa}} e^{- \left(\omega_{\kappa}+\w\right)^2} \kappa^2 \left(1+2 \kappa^2 +2\m^2 +2 \omega_{\kappa}  \w -e^{2 \omega_{\kappa}
\w } \left(1+2 \kappa^2 +2 \m^2\right) \text{sinc}\left(\kappa \l\right)\right),
\end{equation}
which is identically zero at $\m = 0$. In fact, the first derivatives of both $\mathcal{L}$ and $-\text{Re}\,\mathcal{M}$ with respect to the mass are zero at $\m = 0$. This is expected, since all these terms are differentiable at $\mu=0$ and they depend on $\mu$ only through $\mu^2$. The second derivative of $\mathcal{N}^+$ evaluated at $\mu = 0$ is given by
\begin{align}
    \left.\frac{\partial^2 \mathcal{N}^{+}}{\partial \m^2}\right|_{\m = 0} &= \frac{\lambda^2}{4\pi^2} \int_{0}^{\infty}\frac{\dd \kappa}{\kappa} e^{- (\kappa+\w )^2}  \left(1+2 \kappa^2 +2 \kappa \w -e^{2 \kappa  \w } \left(1+2 \kappa^2\right) \text{sinc}\left(\kappa \l\right)\right) \label{eq:secondDerivative}\\
    &= \lambda^2\frac{e^{-\w^2}}{4\pi^2} \left(1+ {}_2\!\:\!F_2\!\left(1,1;2,\tfrac{3}{2};\w^2
    \right)\w^2+ {}_2\!\:\!F_2\!\left(1,1;2,\tfrac{5}{2};-\tfrac{\l^2}{4 }\right)\tfrac{\l^2}{16}\nonumber 
    -\tfrac{\pi}{2}  \text{erfi}(\w)-\tfrac{2}{\l} \mathcal{F}\left(\tfrac{\l}{2 }\right)\right),
\end{align}
where $\mathcal{F}$ is the Dawson function and ${}_pF\!{}_q$ is the generalized Hypergeometric function
\begin{align}
    \mathcal{F}(z) := e^{-z^2} \int_{0}^{z} e^{y^2} \dd y, \quad\quad\quad
    {}_pF\!{}_q(a_1, \ldots a_p;b_1, \ldots, b_q;z) := \sum_{k=0}^{\infty}\frac{(a_1)_k, \ldots (a_p)_k}{(b_1)_k, \ldots, (b_q)_k}\frac{z^k}{k!},
\end{align}
and $(x)_k$ is the Pochammer symbol,
\begin{equation}
    (x)_k:= \frac{\Gamma(x+k)}{\Gamma(x)}.
\end{equation}
Finally, we address the case of spatially smeared detectors. The integrals in Eqs. \eqref{eq:firstDerivative} and \eqref{eq:secondDerivative} are modified by simply introducing the term $e^{-\kappa^2 \d^2}$ in the integrand. Applying the same procedure as above, and expanding in $\sigma$, we find
\begin{align}
    \mathcal{N}^{+}_{\m = 0} =& \frac{\lambda^2e^{-\w^2}}{8\pi^2} \bigg(\sqrt{\pi}\tfrac{ 1}{\l}e^{-\frac{\l^2}{4}} \text{erfi}\left(\l/2\right)+\sqrt{\pi}\w e^{\w^2}\text{erfc}(\w)- 1 \bigg) \nonumber \\
    &+\lambda^2\d^2\frac{e^{-\omega^2}}{16 \pi^2} \left(1+2\w^2+ \left(\l-\tfrac{2}{\l}\right) \mathcal{F}\left(\l/2 \right) - e^{\omega^2}\sqrt{\pi}\w(3+2\w^2)\text{erfc}(\w) \right)+\mathcal{O}(\sigma^3), \\
    \left.\frac{\partial^2 \mathcal{N}^{+}}{\partial \m^2}\right|_{\m=0}=& \lambda^2\frac{e^{-\w^2}}{4\pi^2} \left(1+ {}_2\!\:\!F_2\!\left(1,1;2,\tfrac{3}{2};\w^2
    \right)\w^2+ {}_2\!\:\!F_2\!\left(1,1;2,\tfrac{5}{2};-\tfrac{\l^2}{4 }\right)\tfrac{\l^2}{16}\nonumber 
    -\tfrac{\pi}{2}  \text{erfi}(\w)-\tfrac{2}{\l} \mathcal{F}\left(\tfrac{\l}{2 }\right)\right)\nonumber \\&+ \lambda^2\sigma^2 \frac{1}{4\pi^{2}}\left(\frac{3}{2} \w\sqrt{\pi} \text{erfc}(\w) -e^{-\w^2}\left(1+ \sqrt{\pi}\tfrac{1}{\l}e^{ - \frac{\l^2}{4}}\text{erfi}\left(\l/2 \right) \left(\tfrac{\l^2}{ 4}-1\right)\right)\right)+\mathcal{O}(\sigma^3).
\end{align} 
With the results above, we find an expression for the harvested negativity that has the form
\begin{equation}
    \mathcal{N}^{+}(\w,\l,\m,\d) \approx \mathcal{N}_0^{+}(\w,\l) + \frac{\m^2}{2}\partial_{\m}^2\mathcal{N}_0^{+}(\w,\l) + \frac{\d^2}{2} \partial_{\d}^2\mathcal{N}_0^{+}(\w,\l),
\end{equation}
where we use the subindex $0$ to denote evaluation at $\m = \d = 0$. Using the expression above, it is possible to find an approximate expression for the value of $\w$ that maximizes the negativity, by imposing
\begin{equation}
    \pdv{\mathcal{N}^{+}}{\w} = 0\quad\Rightarrow\quad\pdv{}{\w} \mathcal{N}_0^{+}(\w,\l)+\frac{\m^2}{2}\pdv{}{\w}\left(\partial_{\m}^2\mathcal{N}_0^{+}(\w,\l)\right) + \frac{\d^2}{2} \pdv{}{\w}\left(\partial_{\d}^2\mathcal{N}_0^{+}(\w,\l)\right) = 0.\label{eq:omegal}
\end{equation}
\end{widetext}
Unfortunately, the expression above does not admit a solution in terms of elementary functions. However, as discussed in Subsection \ref{sec:optimizing}, the solution for large $\l$ in the case $\m = \d = 0$ is approximately $\w = \l/2$. This suggests that we can write the solution to Eq. \eqref{eq:omegal} as $\w = \l/2+\epsilon$, where $\epsilon$ is a small parameter that can depend on $\l$, $\m$ and $\d$. Performing an expansion in $\epsilon$ to second order, we obtain a quadratic equation, which can be used to approximate the solution of Eq. \eqref{eq:omegal}. That is, we find $\epsilon(\l,\m,\d)$ such that $\omega = \l/2 + \epsilon$ is an approximate solution of Eq. \eqref{eq:omegal}. Moreover, we can expand $\epsilon(\l,\m,\d)$ for small values of mass and detectors sizes, obtaining a closed form result:
\begin{equation}
    \epsilon(\l,\m,\d) = \mathcal{E}(\l) + \m^2\mathcal{A}(\l) + \d^2 \mathcal{B}(\l) +\mathcal{O}(\d^2\m^2).
\end{equation}
However, the closed expressions for $\mathcal{E}(\l)$, $\mathcal{A}(\l)$ and $\mathcal{B}(\l)$ are too cumbersome to provide us with any insightful intuition. Nevertheless, each of the terms $\mathcal{E}(\l)$, $\mathcal{A}(\l)$ and $\mathcal{B}(\l)$ can be very well approximated by simpler functions of $\l$. We use Mathematica to find best fits for each of these terms. We find that the $\mathcal{E}(\ell)$ function can be well fit by a function of the form $a_1/\ell$, $\mathcal{A}(\ell)$ can be approximately described by $b_1/\ell^{c_1} + a_2$ and $\mathcal{B}(\ell)$ admits a linear fit of the form $b_2 \ell + a_3$ with $a_1 \approx -1.39218$, $a_2\approx -0.0230021$, $a_3\approx 0.377855$, $b_1\approx -0.987746$, $b_2\approx -0.226636$, and $c_1\approx 1.25143$. These estimates provide an $L^1$ relative error smaller than $0.5\%$ for values of $\ell\geq 5$. Thus, the functions $\mathcal{E}(\ell)$, $\mathcal{A}(\ell)$ and $\mathcal{B}(\ell)$ are all negative functions of $\ell$ when the detectors are approximately spacelike separated. This implies that the negativity peaks happen at a frequency that is a little smaller than $\ell/2$, and decreases with the mass of the field and detector separation.


\bibliography{references}
    
\end{document}